\documentclass[a4paper,11pt]{article}
\usepackage{graphicx}
\usepackage{fullpage}
\usepackage{amsmath}
\usepackage{amsfonts}
\usepackage{amssymb}
\usepackage[round]{natbib}
\usepackage{color}
\usepackage[titletoc,toc,title]{appendix}
\usepackage{authblk}
\usepackage{subfigure}

\usepackage{prettyref}
\newrefformat{hypo}{H\ref{#1}}
\newcounter{hypothese}

\newtheorem{theorem}{\sc Theorem}

\def\R{\mathbb R}

\def\bK{\boldsymbol{K}}

\def\bX{\boldsymbol{X}}
\def\bY{\boldsymbol{Y}}

\begin{document}

\title{
Estimation of temperature-dependent growth profiles for the assessment of time of hatching in forensic entomology}

\author{D. Pigoli}\affil{Department of Mathematics, King's College London} \author{J.A.D. Aston}\affil{Statistical Laboratory, DPMMS,
University of Cambridge} \author{F. Ferraty}\affil{Toulouse Mathematics Institute, University of Toulouse} \author{A. Mazumder}\affil{The Alan Turing Institute} \author{C. Richards}\affil{Department of Life Sciences, Natural History Museum, United Kingdom} \author{M.J.R. Hall}\affil{Department of Life Sciences, Natural History Museum, United Kingdom}
\date{}
\maketitle

\begin{abstract}
Forensic entomology contributes important information to crime scene investigations. In this paper, we propose a method to estimate the hatching time of larvae (or maggots) based on their lengths, the temperature profile at the crime scene and experimental data on larval development. This requires the estimation of a time-dependent growth curve from experiments where larvae have been exposed to a relatively small number of constant temperature profiles. Since the temperature influences the developmental speed, a crucial step is the time alignment of the curves at different temperatures. We propose a model for time varying temperature profiles based on the local growth rate estimated from the experimental data. This allows us to estimate the most likely hatching time for a sample of larvae from the crime scene. Asymptotic properties are provided for the estimators of the growth curves and the hatching time. We explore via simulations the robustness of the method to errors in the estimated temperature profile. We also apply the methodology to data from two criminal cases from the United Kingdom.
\end{abstract}

\section{Introduction}\label{s:intro}

Forensic entomology is the study of insects (and other arthropods) in relation to criminal investigation, frequently involving insect evidence in cases of suspicious death. A great diversity of insects are attracted to decomposing human corpses, both to feed and to lay their eggs or larvae. Flies and beetles are the most common visitors, as both immature insects (larvae or maggots) and adults. In particular,  blow flies are among the most important insects in criminal investigation, because they are usually the first to arrive \citep{greenberg1991flies}. While the methods described in this work could be applied in general to any insects of forensic interest,  we will consider here two species of blow fly, \emph{Calliphora vicina} and \emph{Calliphora vomitoria} (Diptera:
Calliphoridae).

Fly larvae are usually used to estimate a lower bound for the time since death, also known as post-mortem interval. Indeed, the age of the oldest insects on the body provides an assessment of when the mother insects gained access to the body and, as a consequence, a lower bound for the post-mortem interval in a criminal investigation. In this work we focus in particular on the information that is possible to obtain from the length of the larvae observed on the body or at the crime scene. The life cycle of flies is divided into different stages. In the case of an outdoor crime scene, adult female blow flies can arrive within just a few hours to lay eggs on the body \citep{reibe2010promptly, hofer2017optimising}. The eggs then hatch into first instar larvae (commonly called maggots) which start feeding and subsequently develop into second and then third instar larvae (between these stages, the cuticle is shed to allow for growth). When larvae have finished feeding, they usually disperse from the body and the cuticle shrinks and hardens, leading to the immobile puparia within which the pupal and pharate adult stages develop through
metamorphosis \citep{martin2017looking}. Finally, adult flies emerge from puparia, feed, mate and begin a new cycle.

\begin{figure}[t!]
	\centering
	\includegraphics[width=0.7\textwidth]{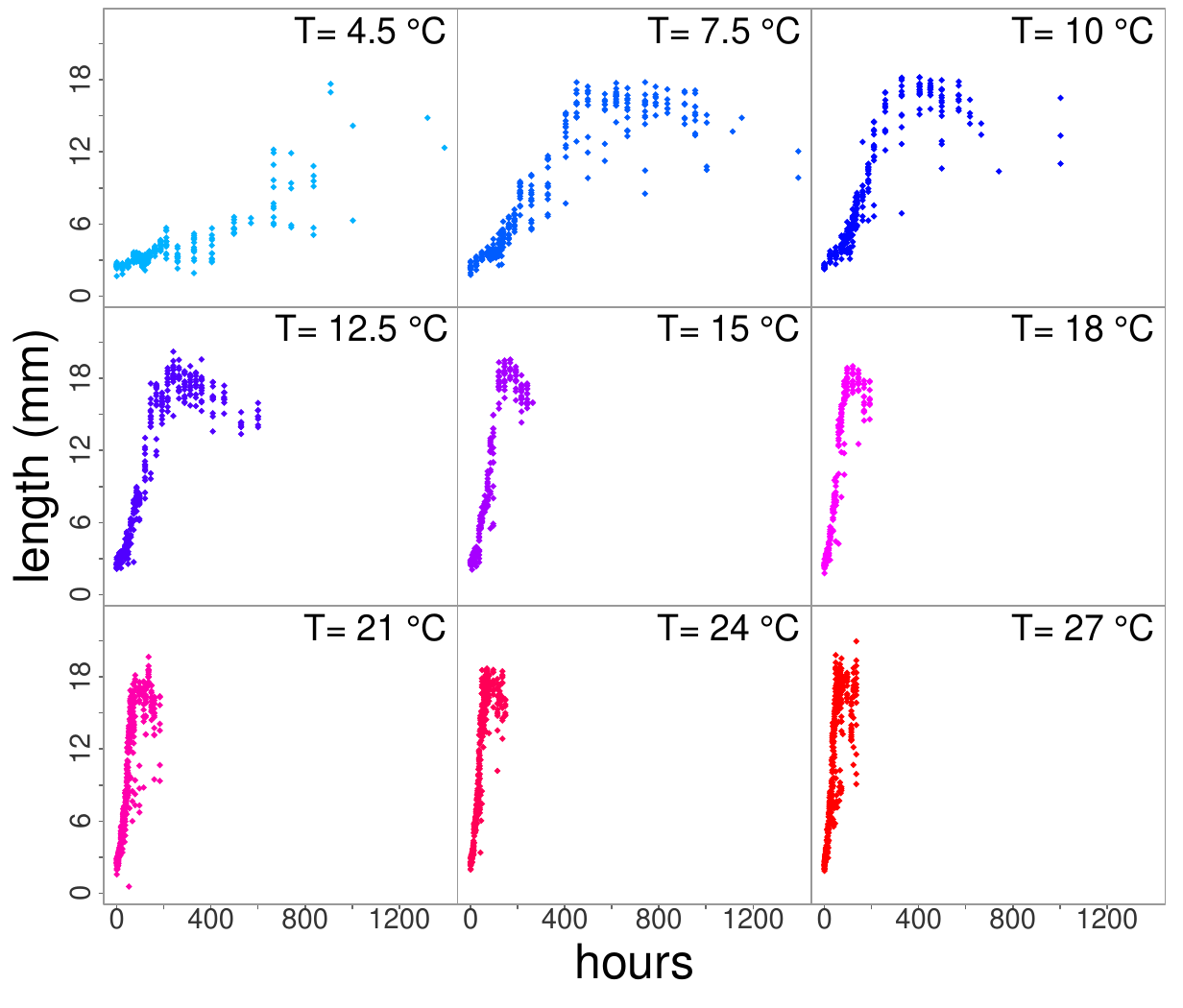}
	\caption{\label{fig2} Experimental development data for \emph{Calliphora vicina}, measured at $9$ different constant temperature profiles. Given the difference in time scale between experimental conditions, different sampling rate were also used, resulting in the total number of measurements of larval lengths being respectively $271$ (T=$4.5^{\circ}$C), $344$ (T=$7.5^{\circ}$C), $275$ (T=$10^{\circ}$C), $351$ (T=$12^{\circ}$C), $261$ (T=$15^{\circ}$C), $224$ (T=$18^{\circ}$C), $483$ (T=$21^{\circ}$C), $449$ (T=$24^{\circ}$C), $437$ (T=$27^{\circ}$C). Further details on the experimental procedure can be found in \citet{donovan2006larval}.}
\end{figure}

The rate of development of fly larvae is temperature dependent and when comparing the specimens observed at the crime scene with experimental data it is essential to adjust outcomes with respect to the different temperatures. Currently, the standard technique used to estimate the rate of development with temperature compensation is the Accumulated Degrees Hours (ADH) model. This consists in summing the hours of development multiplied by the temperature (in $^{\circ}$C), thus providing a rough measure of the amount of thermal energy available to the larva. This can then be compared with existing experimental results about how many ADH are needed to reach each developmental stage in the insect life cycle and from this deduce the time of colonization at the scene \citep{amendt2007best}. Our goal is to provide more refined methods that consider the growth dynamic between hatching and pupariation, mostly ignored by previously existing techniques. We will rely on experimental development data on larval lengths for \emph{Calliphora vicina} (see Figure \ref{fig2}) and \emph{Calliphora vomitoria} \citep{richards2017} to achieve this.

While experimental data are usually collected at constant temperatures, real life scenes
are subjected to dynamic temperature profiles. In this work, we apply techniques from functional
data analysis \citep[FDA;][]{ramsay2005functional,ferraty2006nonparametric,horvath2012inference}  to
reconstruct the growth curve of the larvae in the presence of a varying temperature profile, using
information from experiments run at constant temperatures (see again Figure \ref{fig2}). The objective is the assessment of the time since colonization, which is the interval between the earliest time when the observed blow fly specimens were laid as eggs and the time when they were collected at the scene. The current approach is often to use the ADH model which uses an average temperature approach.  A more accurate assessment of the hatching time for a given set of larvae, via a case specific growth curve, would be a useful addition to the forensic science toolbox. This would be relevant for many criminal investigations, since evidence from forensic entomology is used in tens of high-profile cases every year in the United Kingdom alone, and many more worldwide \citep{tomberlin2015forensic}.

From a statistical perspective, FDA is the field that investigates the statistical properties of a collection of curves, surfaces or any more complex mathematical object. In forensics, FDA methods have already proved useful in analysing data generated from spectroscopic techniques \citep{dias2013hierarchical,burfield2015review}. More recently, FDA has also started to be of interest in forensic entomology, with one preliminary investigation using the technique for thermal wavelength analysis of maggot populations for time of death prediction \citep{WARREN2017205}. Here, we use the well known connections of FDA to growth curve analysis \citep{ramsay2005functional}.  

The main challenge in this paper is the need to combine different sources of information about the growth process to estimate the most likely growth curve that led to the larval lengths observed at the scene. Larval development data are collected in incubators which keep the temperature constant, while larvae at the crime scene are exposed to a varying temperature profile. The first part of the methodology requires an estimate of the actual case specific growth curve from the growth curves observed at constant temperatures in the lab. Then, the hatching time that leads to the best fit of the measured larval lengths at the crime scene is selected. The method has been developed with considerable attention paid to the pieces of information and data that are available in practice to forensic scientists. To support the validity of the proposed estimator, in addition to providing asymptotic results both for the estimator of the growth curve and of the hatching time, we assess the finite sample properties of the procedure via simulation studies. Finally, we discuss the application of the method to the data from two investigations in the United Kingdom. It is also worth mentioning that, while we developed our procedure in view of the application in forensic entomology which is described below, the methods and theory we discuss here for the estimation of growth curves can be relevant in other biological applications, such as microbiology \citep{zwietering1990modeling} or quantitative genetics \citep{kingsolver2004quantitative}.

\section{Model and methods}
The fact that larval development depends mainly on temperature is well known in forensic entomology \citep{donovan2006larval, amendt2007best}. In addition, outdoor temperatures vary with time. However, laboratory experiments are bound to measure this relationship only for a relatively small set of temperatures, as shown in Figure \ref{fig2}. For each plot, the growth process is observed for a constant temperature profile (i.e. same temperature from hatching to pupariation). Our first goal is to use techniques from functional data analysis and nonparametric regression to estimate the expected growth curve corresponding to realistic temperature profiles. Once growth lengths are available for any temperature profile, one is able to estimate the hatching time and hence the interval between hatching and body discovery from data available at the scene.

\subsection{Estimation of the growth process}\label{s:growth}
\noindent{\em Experimental larval development data}. In a typical larval development data set, we have $K$ experimental temperatures $T_1<T_2<\cdots<T_K$ and, for each species of interest, one observes the larval lengths $Y_{kjl}$ measured at time $t^k_j$ after hatching for $j=1,\ldots,n_k$ and for $l=1,\ldots,N_{kj}$ individual larvae at each time point which has been exposed to a constant experimental temperature $T_k$, with $k=1,\ldots,K$. The observation times $t^k_1,\ldots,t^k_{n_k}$ may differ across experimental temperatures, generally being set at wider intervals for lower temperatures and as a result of the experimental procedure a different number of individual larvae may be measured at each observation time (see for example \citet{donovan2006larval} for more details). Figure \ref{fig2} illustrates an example of this kind of experimental data for \emph{Calliphora vicina}. In this example, we have $K=9$ values of temperature set in the incubator, for each value of temperature $T_k$ we have a number of observation times $n_k$ which is between $16$ (for $T_6=18^{\circ}$C) and $34$ (for $T_9=27^{\circ}$C) and for most of observation times $t_j^k$ we have about $15$ measured larval lengths (fewer measurement are available for later times, as can be seen from Figure \ref{fig2}).  In these data, we conventionally set the hatching time $t_h$ to be zero (i.e. $t_h = t_1 = 0$) so that it is the time reference.\\ 

\begin{figure}[h]
	\centering \includegraphics[width=0.55\textwidth]{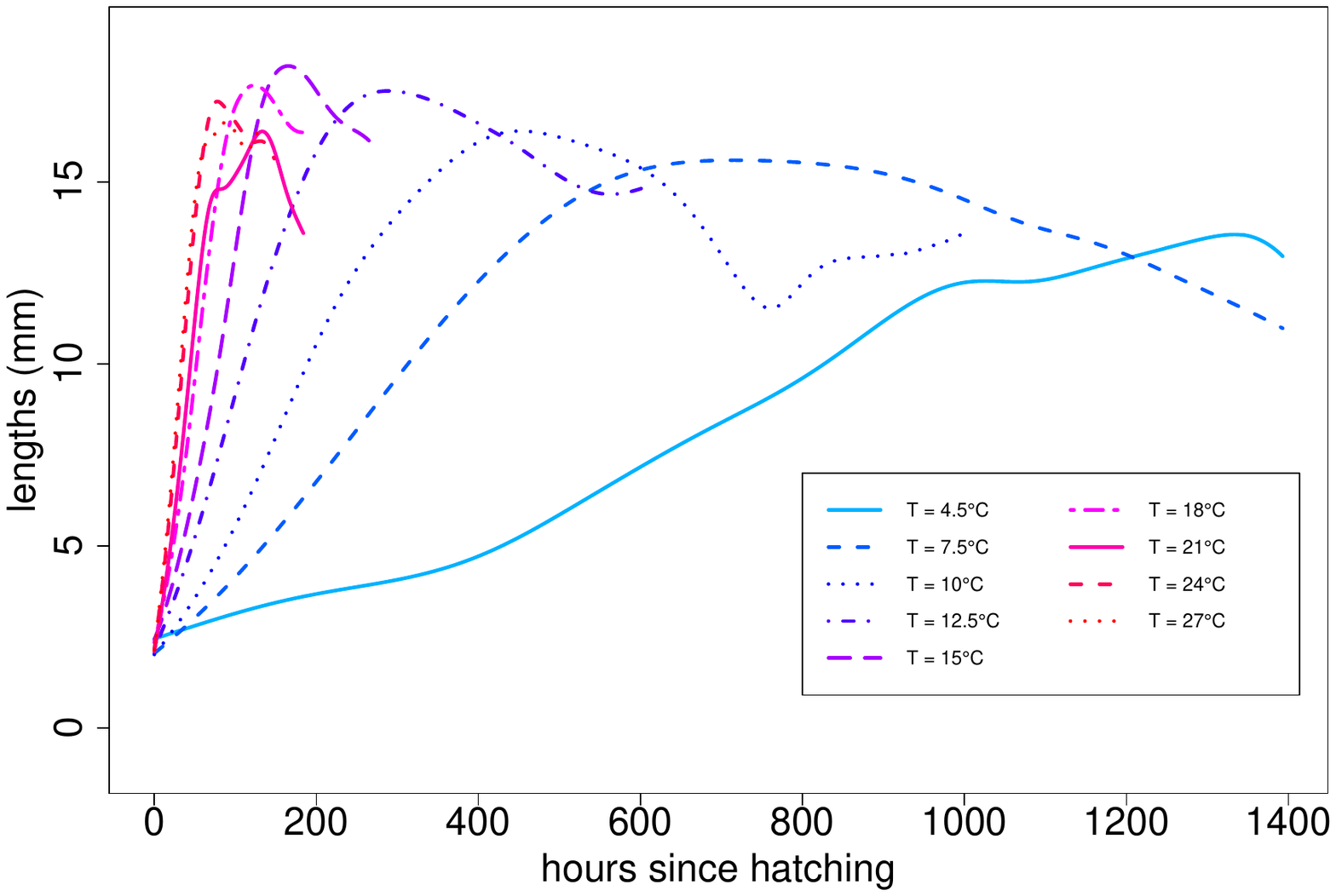}
	\caption{\label{fig3} Smooth experimental growth curves estimated from the \emph{Calliphora vicina} lengths measured at constant temperature $T$. Clearly the larvae at $4^{\circ}$C developed with a different profile to those at all higher temperatures and none were observed to pupariate, most probably because $4^{\circ}$C was near to the estimated lower development threshold for this species of $1.0^{\circ}$C-$1.5^{\circ}$C \citep{donovan2006larval}.}
\end{figure}

\noindent{\em Estimating mean larval length curves for any experimental temperatures and their derivative}. We can then assume that the observed lengths satisfy a nonparametric regression model $Y_{kjl}=L_{T_k}(t_j)+\epsilon_{kjl}$, where $\epsilon_{kjl}$ are independent, zero mean random variables and the mean larval length curve $L_{T_k}$ depends on the experimental temperatures $T_k$, $k=1,\ldots,K$. It is then possible to estimate $L_{T_k}$ by means of a nonparametric smoothing estimator. In this work, we use a local linear regression estimator (see for instance \citealt{fan1992}, \citealt{fan1992variable}, \citealt{fan1993local}, and \citealt{ruppert1994multivariate}). It is well known in the nonparametric statistical literature that local linear regression provides accurate estimates $\widetilde{L}_{T_k}$  for all growth curves observed at a constant temperature. However, as explained in the next paragraph, the introduced dynamic model for determining the hatching time requires the estimation of the derivative $\widetilde{L}'_{T_k}$ of $\widetilde{L}_{T_k}$ and hence local linear regression is particularly suited to our setting. Figure \ref{fig3} displays the estimated growth curves $\widetilde{L}_{T_1},\ldots,\widetilde{L}_{T_{K}}$ at any time $t$ in their respective time ranges for the \emph{Calliphora vicina} data.\\

\noindent{\em Differential equation model}. The question now is how to produce an estimate for the growth curve at a grid of time points $t_1, \ldots, t_M$ associated to a generic temperature profile $T(t_1),\ldots, T(t_M)$. We claim that, if we consider a small enough time interval, the temperature in that interval can be considered roughly constant and therefore the growth process would be bound to follow the local dynamics of the correspondent constant temperature growth curve \emph{at the corresponding stage of the growth process}, i.e. at the point of the curve which reaches the current length in the growth process. This suggests the following differential equation model (DEM) for the local growth process:

\begin{equation}\label{eq_growth_model}
	\left.\frac{d\,L(t)}{dt}\right|_{t=t_m}\, =
	\left.\frac{d\,L_{T(t_m)}(u)}{du}\right|_{u=L_{T(t_m)}^{-1}\left(L(t_m)\right)},
\end{equation}
where
$L_{T(t_m)}$ is the growth profile at constant temperature $T(t_m)$ and $L(t)$ is the growth profile with varying temperature. The differential equation model (\ref{eq_growth_model}) can be reformulated as $L'(t) = L'_{T(t_m)}  \circ L^{-1}_{T(t_m)} \circ L(t_m)$ where $L'$ stands for the derivative of $L$ and the symbol $\circ$ refers to the usual composition operator between real-valued functions.
This dynamic model means that the (expected) local increment in length at time $t_m$ is the one that would occur in the growth profile at constant temperature $T(t_m)$ when the length is equal to $L(t_m)$. This allows us to reconstruct the varying temperature growth profile iteratively, as a discretised solution to the above DEM:

$$
\left\{
\begin{array}{l}
	L(t_1)=L_{T(t_1)}(t_1),\, \mathrm{and\, for}\, m=1,\ldots,M-1,\\
	L(t_{m+1})=L(t_m)+ (t_{m+1}-t_m) \left\{L'_{T(t_m)} \circ L^{-1}_{T(t_m)} \circ L(t_m)\right\},
\end{array}
\right.
$$
with $T(t_1),\ldots T(t_M)$ being the varying temperature profile. 

This would solve the problem if we knew the expected growth curve $L_T$ and its derivative $L'_T$ for any temperature $T$ we can observe, but in practice we can have experimental data only for a relatively small set of temperatures. \\

\begin{figure}[t!]
	\centering \makebox{\includegraphics[scale=0.3]{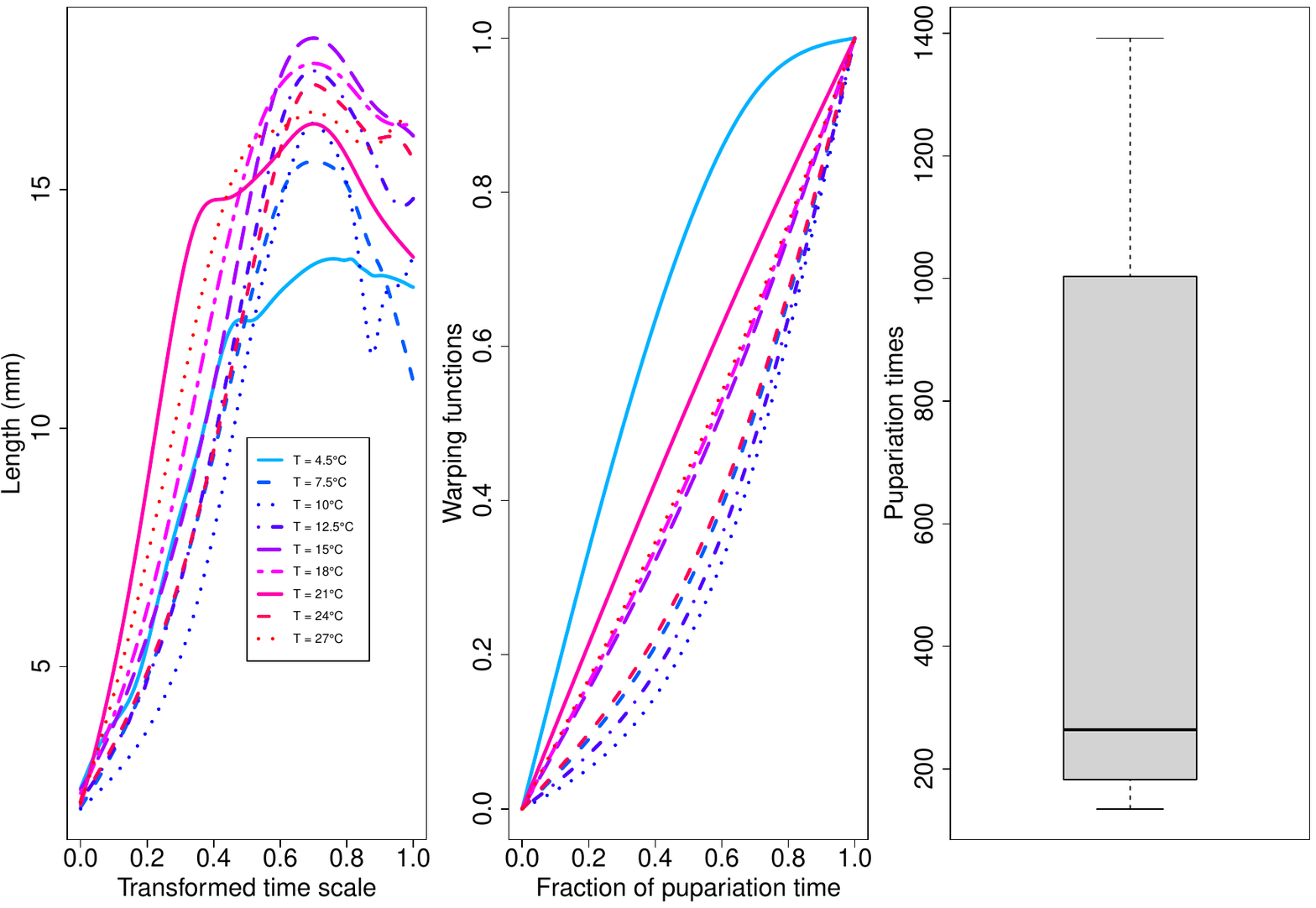}}
	\caption{\label{fig4} Growth shapes $\widetilde{S}_{T_1}, \ldots, \widetilde{S}_{T_K}$ (left), warping functions $ \widetilde{w}_{T_1}, \ldots, \widetilde{w}_{T_K}$ (center) and boxplot of the puparation times (right) for the growth curves $\widetilde{L}_{T_1}, \ldots, \widetilde{L}_{T_K}$ estimated from the \emph{Calliphora vicina} experimental larval developmental data.}
\end{figure}

\noindent {\em Estimating growth profile and its derivative for any temperature}. We need to estimate first the growth profile $L_T$ for a generic constant temperature $T$ from a set of estimated growth curves $ \widetilde{L}_{T_1}, \ldots, \widetilde{L}_{T_K}$. The main difficulty here is that the temperature influences the speed of the growth process. Using the language of functional data analysis, these curves present both amplitude and phase variation \citep[see][]{ramsay2005functional,marron2014statistics}, as can be appreciated from the example of the \emph{Calliphora vicina} experimental data and their estimated growth curves at constant temperature (see Figure \ref{fig3}). At this stage, one postulates that the mean larval length curves  $L_{T_1},\ldots,L_{T_{K}}$ have corresponding profiles in a standardized time scale $S_{T_1},\ldots,S_{T_K}$. These can be obtained using warping transformations $w_{T_1},\ldots,w_{T_K}$ acting over the time in such a way, for any $k=1,\ldots,K$ and for all $t$ in $[0, t_{pup,T_k}]$,  $S_{T_k}\circ w_{T_k}(t/t_{pup,T_k})=L_{T_k}(t)$, where $t_{pup,T_k}$ is the pupariation time (the last time in the experiment when the lengths of larval specimens were recorded) at constant temperature $T_k$. The warping functions are assumed to be one-to-one nondecreasing functions mapping $[0,1]$ into $[0,1]$ and restricted to the space of quadratic polynomials. They can be determined via a landmark registration procedure that aligns a set of structural points featuring the sample of curves \citep[see, e.g.,][Chapter 7]{ramsay2005functional}. In our application, hatching time, time of maximum length and pupariation time are three structural points that correctly depict the growth profiles. We propose to match all growth profiles with respect to these three structural landmarks. The procedure achieving the alignment of a collection of profiles is well known in the functional data analysis literature under the terminology {\em curves registration} \citep[see][]{kneip1992statistical,  ramsay1998curve, maldonado2002similarity, wang1997alignment}. It is of course possible to resort to more recent registration methods \citep[see, e.g.,][]{james2007curve, kneip2008combining, srivastava2011registration} but the relatively simple shape of the growth curves allows landmark registration to provide easily and quickly two collections of simple curves: $K$ estimated growth shapes $\widetilde{S}_{T_1},\ldots,\widetilde{S}_{T_K}$ (i.e. the registered curves) and $K$ estimated warping functions $\widetilde{w}_{T_1},\ldots,\widetilde{w}_{T_K}$ (see Figure \ref{fig4}). From a practical point of view, the warping functions $\widetilde{w}_{T_k}$'s are first computed (see {\em Step 2} in Section \ref{sec:asymp} and the proof of {\sc Theorem} \ref{th2} for their exact definitions) and then the growth shapes $\widetilde{S}_{T_k}$'s are derived with the decomposition resulting from the registration procedure: for any $u\in[0,1]$, $\widetilde{S}_{T_k}(u) = \widetilde{L}_{T_k}\left( t_{pup, T_k} \widetilde{w}^{-1}_{T_k}(u) \right)$.

The growth shape and warping function are reconstructed at any temperature $T$ by function-on-scalar nonparametric regression (see \citealt{ferraty2011kernel}) as
\begin{equation} \label{eq:shapeatanytemp}
	\widehat{S}_{T}(u)\ =\
	\frac{\sum_{k=1}^{K} \widetilde{S}_{T_k} (u) K_S(h_S^{-1}(T_k-T))}{\sum_{k=1}^{K}  K_S(h_S^{-1}(T_k-T))}
\end{equation}
and
\begin{equation} \label{eq:warpingatanytemp}
	\widehat{w}_{T}(u)\ =\
	\frac{\sum_{k=1}^{K} \widetilde{w}_{T_k}(u) K_w(h_w^{-1}(T_k-T))}{\sum_{k=1}^{K}  K_w(h_w^{-1}(T_k-T))},
\end{equation}
for all $u$ in $[0,1]$, where $K_S$ and $K_w$ are kernel functions and $h_S$ and $h_w$ suitably chosen bandwidths. The nonparametric kernel estimator is especially well adapted for the warping function setting. As a convex combination of the $\widetilde{w}_{T_k}$'s, the estimator $\widehat{w}_{T}$ inherits the properties of the warping functions (i.e. $\widehat{w}_{T}$ is nondecreasing with $\widehat{w}_{T}(0)=0$, $\widehat{w}_{T}(1)=1$).

The estimator of the growth process $L_T$ at any constant temperature $T$ and for all $t$ in $[0,\widehat{t}_{pup,T}]$ is $\widehat{L}_{T}(t)=\widehat{S}_{T} \circ \widehat{w}_{T}(t/\widehat{t}_{pup,T})$, where $\widehat{t}_{pup,T}$ is the predicted pupariation time depending on the temperature $T$. Given the sample of temperatures and corresponding pupariation time $(T_1, t_{pup,T_1}), \ldots, (T_K, t_{pup,T_K})$, the pupariation time for a generic temperature $T$ can be obtained with standard nonparametric regression as
\begin{equation} \label{eq:puptemp}
	\widehat{t}_{pup,T}\ =\
	\frac{\sum_{k=1}^{K} t_{pup,T_k} K_{pup}(h_{pup}^{-1}(T_k-T))}{\sum_{k=1}^{K}  K_{pup}(h_{pup}^{-1}(T_k-T))},
\end{equation}
where $K_{pup}$ is a kernel function and $h_{pup}$ a data driven smoothing parameter. From now on, we are able to estimate the growth profile $L_T(t)$ for any temperature $T$ and for all $t$ in $[0, \widehat{t}_{pup,T}]$.

According to the differential equation model (\ref{eq_growth_model}), we also need to estimate the derivative $L_T'$ of the growth curve $L_T$. We know that for any $t$ in $[0,\widehat{t}_{pup,T}]$, $\widehat{L}_{T}(t)=\widehat{S}_{T} \, \circ \, \widehat{w}_{T}(t/\widehat{t}_{pup,T})$, which results in $\widehat{L}'_{T}(t)=\widehat{S}'_{T} \, \circ \, \widehat{w}_{T}(t/\widehat{t}_{pup,T}) \,  \widehat{w}'_{T}(t/\widehat{t}_{pup,T}) \, \widehat{t}^{-1}_{pup,T}$. The previous steps provide $\widehat{w}_{T}$,  $\widehat{w}'_{T}$ and $\widehat{t}_{pup,T}$. The only quantity we need to recover $\widehat{L}'_{T}$ is $\widehat{S}'_{T}$. For any $u$ in $[0,1]$ and any $k=1,\ldots,K$,  we are able to compute $\widetilde{S}'_{T_k}(u) = t_{pup, T_k} \widetilde{L}'_{T_k}\left( t_{pup, T_k} \widetilde{w}^{-1}_{T_k}(u) \right)  (\widetilde{w}^{-1}_{T_k})'(u)$ since the smoothing step provides $\widetilde{L}'_{T_1},\ldots, \widetilde{L}'_{T_K}$ and the registration procedure gives the exact analytical writing of the warping functions $\widetilde{w}_{T_1}, \ldots,\widetilde{w}_{T_K}$ which enables the calculation of the inverse functions $\widetilde{w}^{-1}_{T_1}, \ldots,\widetilde{w}^{-1}_{T_K}$ as well as their derivatives $\widetilde{w}^{-1'}_{T_1}, \ldots,\widetilde{w}^{-1'}_{T_K}$. Given the sample of temperatures and corresponding growth shapes \\ 
\noindent $(T_1, \widetilde{S}'_{T_1}), \ldots, (T_K, \widetilde{S}'_{T_K})$, the computation of  $\widehat{S}'_{T}$ for any $T$ can be derived from the function-on-scalar nonparametric regression. Thus, one can deduce $\widehat{L}'_{T}$ for any temperature $T$.

~\\ \noindent {\em Conclusion and complement}. The use of our DEM combined with the estimations of $\widehat{L}_T$ and $\widehat{L}'_T$ for any temperature $T$ enable the reconstruction of the growth curve for any temperature profile. The estimating procedure can be divided into four steps: (1) estimation of the mean larval length curves $\widetilde{L}_{T_1},\ldots, \widetilde{L}_{T_K}$ and their derivatives $\widetilde{L}'_{T_1},\ldots, \widetilde{L}'_{T_K}$ for each experimental temperature $T_1,\ldots,T_K$, (2) decomposition of the mean larval length curves into growth shapes and warping functions, (3) estimation of the growth length $\widehat{L}_{T}$ and its derivative $\widehat{L}'_{T}$ at any temperature $T$, (4) for a given temperature profile $\{T(t); \, t\in \mathcal{T}\}$, computation of the corresponding length profile $\{\widehat{L}(t); \, t\in \mathcal{T}\}$ thanks to DEM (\ref{eq_growth_model}). Asymptotic properties for all these estimators are described in Section \ref{sec:asymp} and stated with proofs in the Supplementary Material.

However, a few technical issues need to be considered. First, as the growth curve is not monotone, there may be multiple times $t$ at which $L_{T(t_k)}(t)=L(t_k)$. In particular, we need to distinguish between the feeding phase (initial monotone increase in length up to the maximum) and the post-feeding phase (the usually decreasing region after the maximum). To do this, as long as $\max_{l \leq k} L(t_l) < \max_t L_{T(t_k)}(t)$ the process is labelled as increasing and the positive value of $L'_{T(t_k)} \circ L^{-1}_{T(t_k)} \circ L(t_k)$ is used to obtain the length at $t_{k+1}$. Otherwise, the process is considered as post-feeding and the point corresponding to the negative derivative is used to update the length. Moreover, once the process is in the post-feeding phase, if the length reaches the minimum post-maximum value for the current temperature, i.e. if $L(t_k)\leq \widehat{S}_{T(t_k)}(1)
$, we assume that the process gets to pupariation and the larval length is not defined from that time on.

\subsection{Estimation of the hatching time}
We now want to use the estimated temperature-dependent growth profile to select the most likely
hatching date given a set of length measurements taken at a time $t^*$ where the reference time is the local one. To do that, we compare the growth profiles that would be expected if the hatching time $t_h$ was at any time between the last time the victim has been seen alive $t_a$ and $t^*$. Let $L(t - t_h)$, $t_h\leq
t\leq  t^*$, be the  growth curve for hatching time equal to $t_h$ and temperature profile $\left\{ T(t^* - t); \, t_h \leq t \leq t^* \right\}$ .
Let then $Y^*_{i}$, $=1,\ldots,n_{obs}$ be the measured larval lengths
\begin{equation}
	\label{eq:obs_model}
	Y^*_i=L(t^* - t_h)+\epsilon_i,
\end{equation}
with $\epsilon_i$, $i=1,\ldots, n_{obs}$ independent random errors with zero mean and unknown variance $\sigma^2$. Then, we can estimate $t_h$ as
\begin{equation}
	\label{eq:crit}
	\widehat{t}_h=\arg \min_{t_a \, \leq \, t \, \leq \, t^*} \sum_{i=1}^{n_{obs}}
	\left(\widehat{L}(t^* - t) -
	Y_i\right)^2 = \arg \min_{t_a \, \leq \, t \, \leq \, t^*} \left(\widehat{L}(t^* - t) -
	\overline{Y}\right)^2,
\end{equation}
i.e. we choose the hatching time whose expected length at time $t^*$ best fits the observed values. However, we may also want to include some expert knowledge in the estimation procedure. First, the forensic entomologist collecting the sample may recognise whether the larvae reached the post-feeding phase, i.e. the region of the growth curve after the peak where larvae stop feeding in preparation for pupariation and subsequently decrease in length, or not. We can easily integrate this piece of information in the estimation procedure by restricting the admissible region for the minimisation problem  (\ref{eq:crit}) to the hatching times whose associated growth process at time $t^*$ has already had (or had not) reached the postfeeding region, i.e. the estimated derivative is negative at some time $t\leq t^*$. If we are willing to assume a parametric model for the error, we can also build an approximate confidence interval for the hatching time, for example by inverting the region of non rejection of the likelihood ratio test, i.e. the confidence region will be $CR(\alpha)=\{t:l(t)>l(\widehat{t}_h)-\chi^2_{1-\alpha}(1)/2\}$, where $l$ denotes the log-likelihood and $\chi^2_{1-\alpha}(1)$ the $1-\alpha$ quantile of the $\chi^2$ distribution with $1$ degree of freedom. However, this region is not necessarily convex and we may prefer a more conservative interval defined as $CI(\alpha)=[\min CR(\alpha), \max CR(\alpha)]$. This is a connected interval that, by construction, asymptotically guarantees at least a $(1-\alpha)\%$ coverage of the hatching time. To be coherent with the estimation procedure above, we also need to account for the information about the developmental stage of the observed larvae. Since we optimised the criterion only in the admissible region of hatching time which guarantees the correct developmental stage (postfeeding or not) at the time of sample collection, we need to do the same for the parameter space where the log-likelihood is defined. This is then restricted to the same set of hatching times used in the estimation.

On the other hand, one may want to include prior information about the hatching time, coming for example from the investigative activity (in addition to the interval of admissible hatching times). Let us assume we can translate this
information into a prior distribution on the parameter $t_h$, so that $
t_h \sim \pi$ where $\pi$ is a known distribution. We can then use Bayes
theorem to update the information about $t_h$ given the observed larval lengths and derive
a posterior distribution for $t_h$. For example, assuming a normal distribution for the errors we have
\begin{equation}
	\label{eq:post}
	f(t_h|Y_i)\propto
	\exp\left(-\sum_{i=1}^{n_{obs}}
	\frac{ \{Y_i-L(t^* - t_h)\}^2}{2\sigma^2}\right)
	\pi(t_h).
\end{equation}

In practice we need to substitute $\sigma^2$ with a plug-in
estimate, for example the sample variance of the observed lengths and $L(t^* - t_h)$ with the estimated quantity $\widehat{L}(t^* - t_h)$. Note that in the estimation procedure in this section, we are ignoring the uncertainty in the estimation of the time-dependent growth process. This is indeed present and, while the uncertainty from the experimental data is usually negligible, errors on the temperature at the crime scene can affect the results.  We are going to explore this issue through simulation studies in Section \ref{sec:sim}.
Note that it is straightforward to generalise these ideas to the case where more than one species of larvae (for which developmental data are available) are observed. Let $Y^*_{ij}$ be the observed length of the $i$-th sample from the $j$-th species, $j=1,\ldots,J$,
$$
Y^*_{ij}=L^{(j)}(t^* - t_h)+\epsilon_{ij},
$$
with $\epsilon_{ij}$, $i=1,\ldots, n_{j}$ independent random errors with zero mean and unknown variance $\sigma_j^2$. Then,
\begin{equation}
	\label{eq:mcrit}
	\widehat{t}_h=\arg \min_{t_a \, \leq \, t \, \leq \, t^*} \sum_{j=1}^{J}\frac{\sum_{i=1}^{n_{j}}
		\left(\widehat{L}^{(j)}(t^* - t) -
		Y^*_{ij}\right)^2}{\widehat{\sigma}^2_j/n_j} ,
\end{equation}
where $\widehat{\sigma}^2_j$ is the sample variance of the lengths of the $j$-th species.

To support the validity of the proposed estimator in practice, we investigate its properties in three different ways. Simulation studies and applications of the method to a couple of police case studies can be found in Section \ref{sec:empirics}, while Section \ref{sec:asymp} is devoted to the exploration of the theoretical properties of the estimator in the asymptotic regime. In particular, it is shown in Theorem 4 that the rate of convergence of the estimated hatching time $\widehat{t}_h$ especially depends on the number of larval lengths observed at the crime scene, the grid size where the temperature profile at the crime scene is sampled and the number of experimental temperatures. These quantities have to be as large as possible to ensure an accurate estimation of $\widehat{t}_h$.

\section{Empirical demonstrations}
\label{sec:empirics}
In this section, we first consider some simulation studies to assess the robustness of the proposed method to errors in the temperature profile measured at the crime scene. Then, we present the application of the method to the data coming from two investigations. For reasons of privacy, the cases have
been anonymised and we conventionally set the time at which the measurements of larval lengths have been taken to
be $t^*=0$, with negative times indicating the hours before this moment. In both these forensic cases, we do not have an external corroboration (such as a defendant confession) of the time the body has been abandoned, therefore we compare the results of our procedure with those provided by the ADH method currently used in criminal cases.

Throughout this section, we use quartic kernels in the estimation of the growth shape, the warping function and the pupation time, with the same bandwidth $h=4^{\circ} C$. This is chosen by visual comparison between the estimated growth curves and the observed experimental curves, as it can be seen in Figure \ref{fig:temp303} and \ref{fig:temp307} for the growth profiles of the real criminal case studies. 

\subsection{Simulation studies}
\label{sec:sim} We illustrate here the performance of the proposed methods. We first simulate from model (\ref{eq:obs_model}) using two different temperature profiles. The first, scenario (a), is a constant temperature profile set to $10^{\circ}$C (i.e for all $t$, $T(t)\equiv 10$). The second, scenario (b), corresponds to a more realistic varying temperature profile which is a subset of the temperature measurements from the weather station from the second case study in Section \ref{sec:case2}.

\begin{figure}[h!]
	\centering
	\begin{tabular}{ccc} Scenario (a) & Scenario (b)\\
		\includegraphics[width=0.48\textwidth]{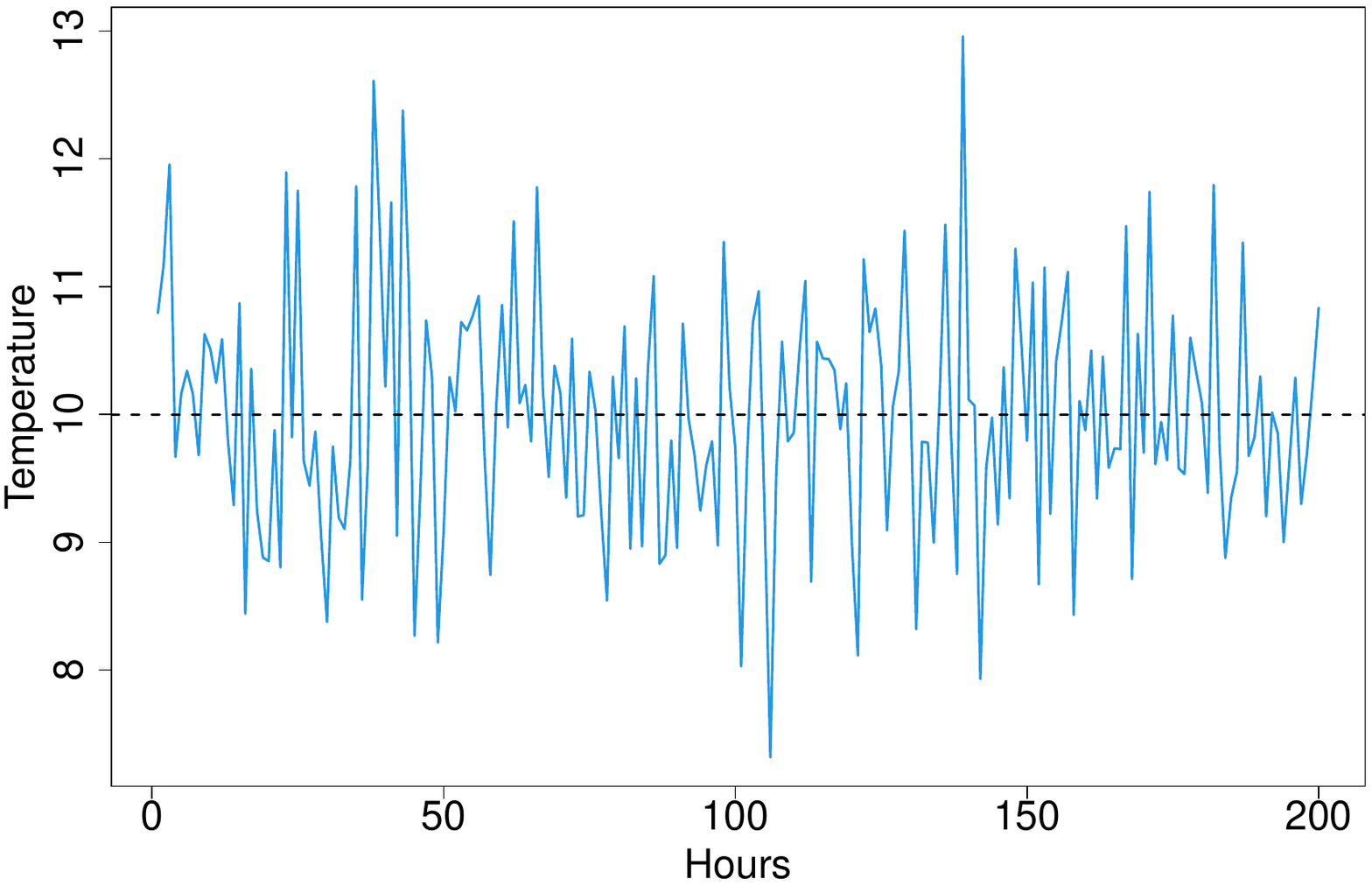} & 
		\includegraphics[width=0.48\textwidth]{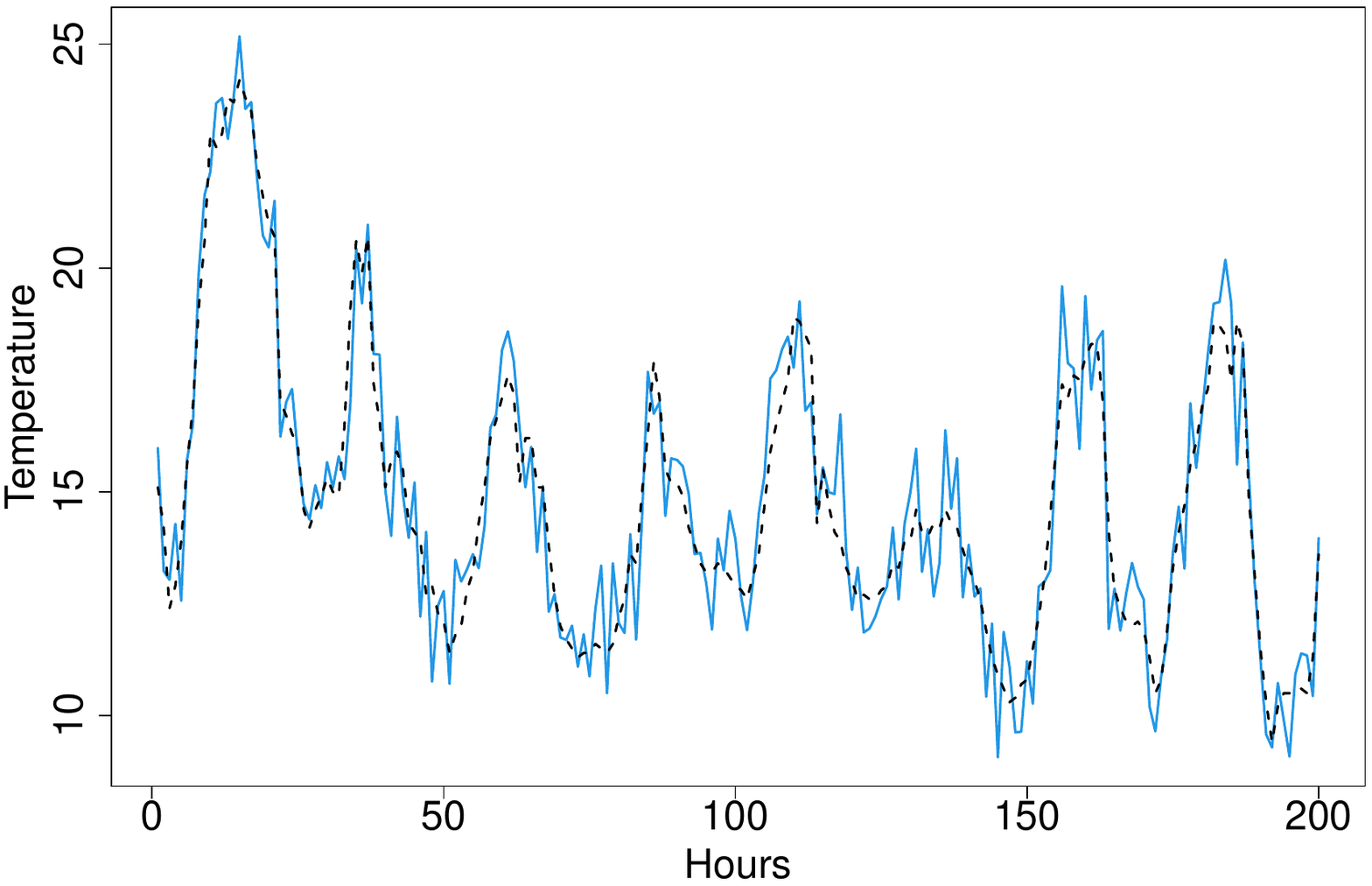} \\
		\includegraphics[width=0.48\textwidth]{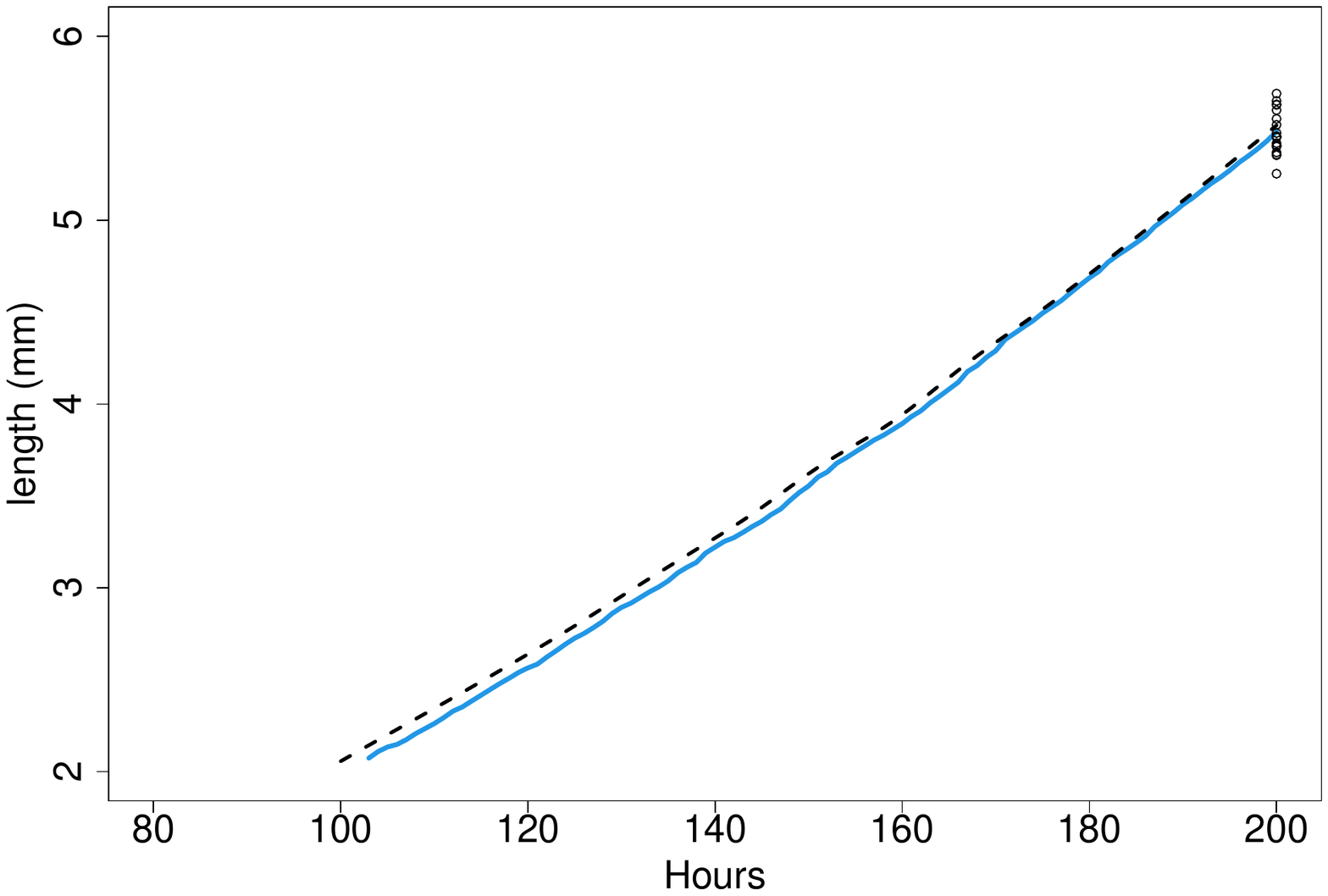} & 
		\includegraphics[width=0.48\textwidth]{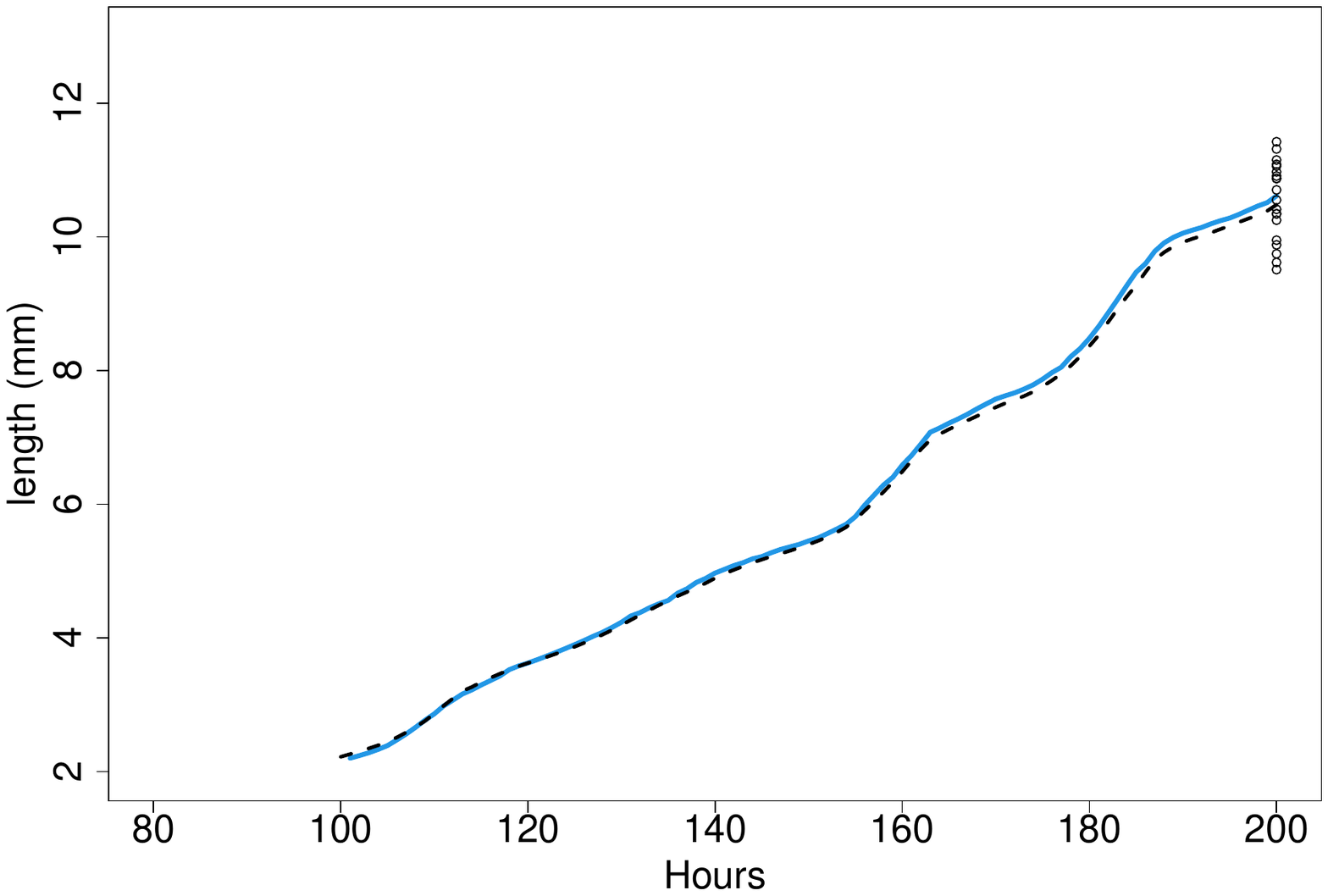}
	\end{tabular}
	\caption{\label{fig:data} Top: Perturbed temperature profile $\widetilde{T}(t_1),\ldots, \widetilde{T}(t_{201})$ with $\sigma_{T}=1$. Bottom: True (black dash line) and estimated (blue solid line) temperature-dependent growth curves with observed larval lengths at the crime scene (black dots). }
\end{figure}

In both cases, an hourly time grid $t_1=-200,\ldots,t_{201}=0$ is used to evaluate the growth curves and optimise the criterion (\ref{eq:crit}) where true hatching time $t_h$ is set to $-100$. We also assess how the results are sensitive to variations in temperatures. To this end, noisy temperatures are build with an independent Gaussian error: $\widetilde{T}(t_k)=T(t_k)+\epsilon_k$ for all $t_k$ in $\{-200,\ldots,0\}$, where $\epsilon_k\sim N(0, \sigma_T^2)$. We consider four different values for $\sigma_T= 0.1,0.25,0.75,1$. For each value of $\sigma_T$, we simulate $1000$ samples of $20$ observed lengths and we estimate 1000 hatching times. Figure \ref{fig:data} give an idea on the shape of temperature profiles under scenario (a) and (b), respectively, and of the corresponding estimated temperature-dependent growth curve. 

\begin{figure}[h!]
	\centering
	\includegraphics[width=0.85\textwidth]{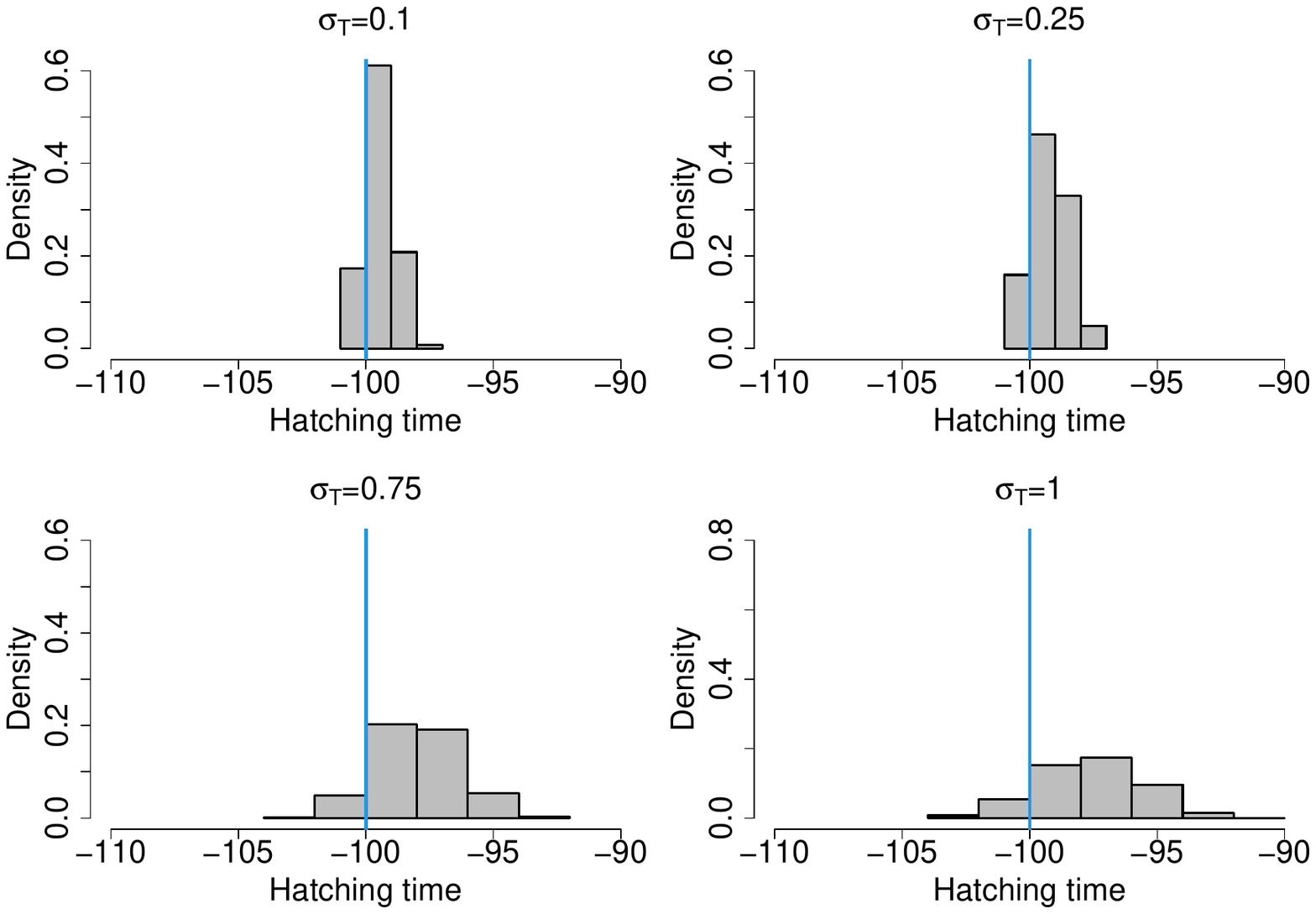}
	\caption{\label{fig:sim1} Scenario (a) - Histograms of the estimated hatching time over $1000$ simulations for different standard deviations $\sigma_{T}$.}
\end{figure}

\begin{figure}[h!]
	\centering
	\includegraphics[width=0.85\textwidth]{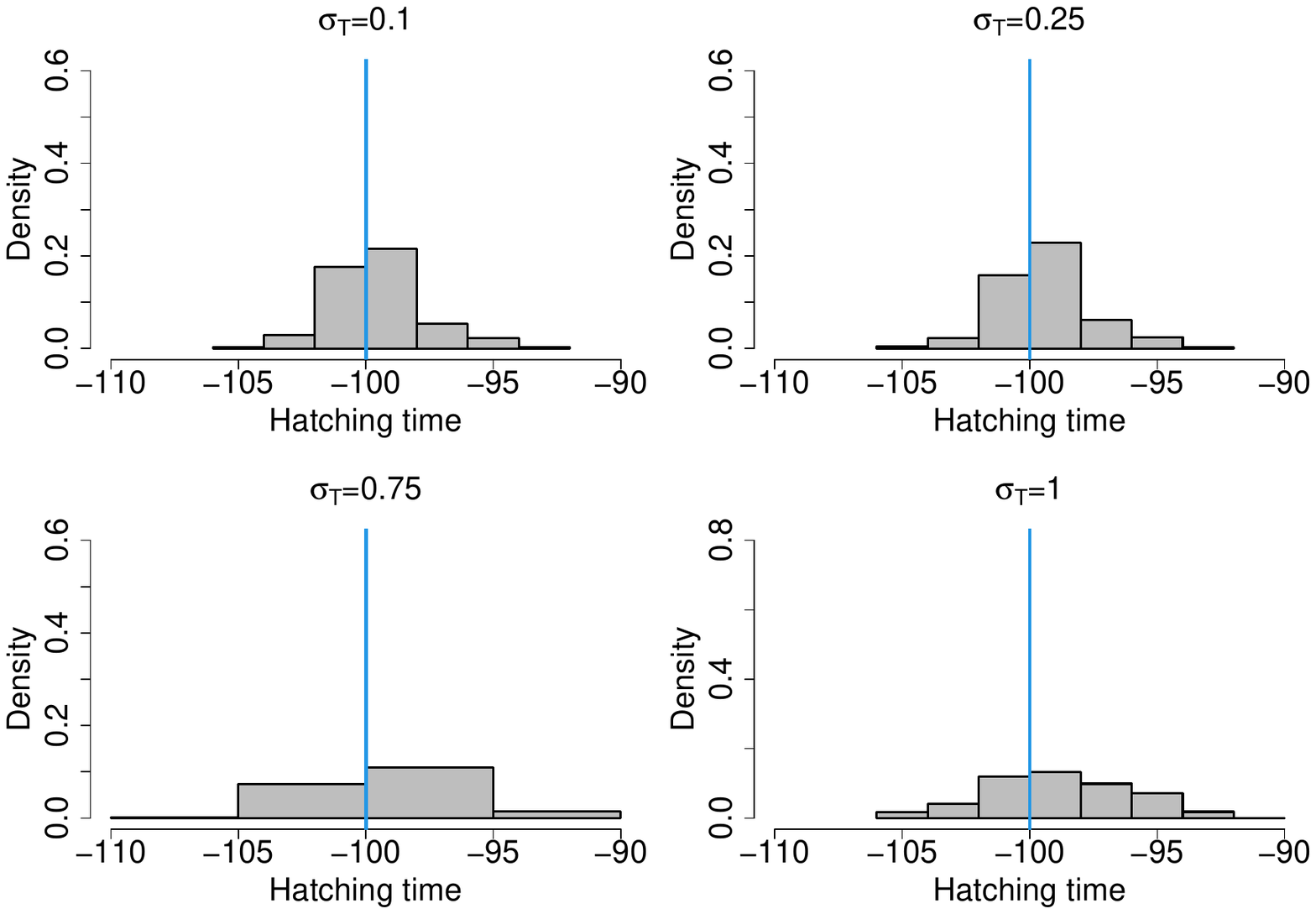}
	\caption{\label{fig:sim3} Scenario (b) - Histograms of the estimated hatching time over $1000$ simulations for different standard deviations $\sigma_{T}$.}
\end{figure}

Figures \ref{fig:sim1} and \ref{fig:sim3} show the distributions of the estimated hatching times for the different values of $\sigma_T$.  The dispersion of the estimates is smaller for scenario (a) which is in accordance with the basic structure of the constant temperature profile.  The variability of the estimates also increases with the variance of the error added to the temperature (i.e $\sigma_T$).

\begin{figure}[h!]
	\centering
	\subfigure[t][]{
		\includegraphics[width=0.45\textwidth]{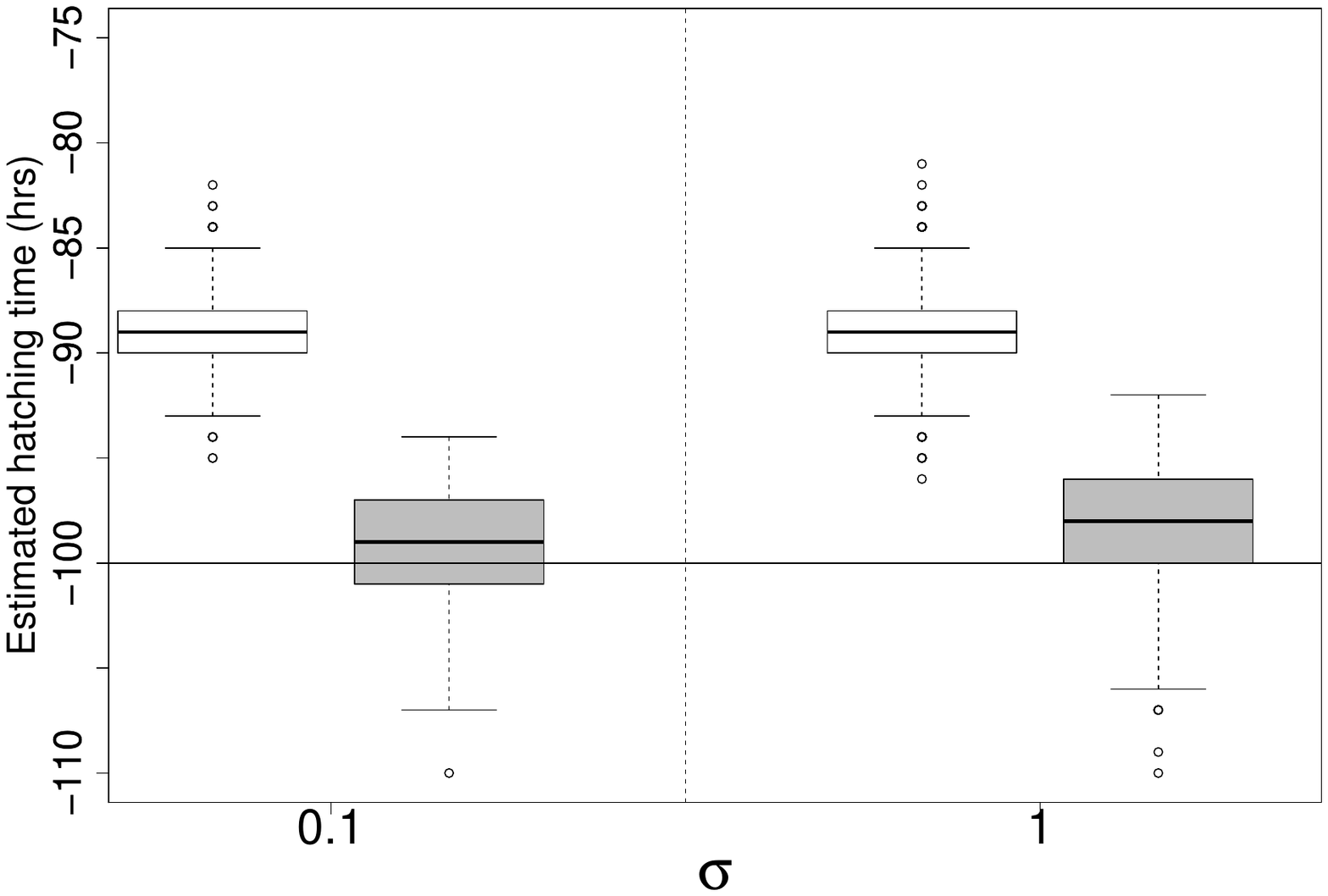}
	}
	\subfigure[t][]{
		\includegraphics[width=0.45\textwidth]{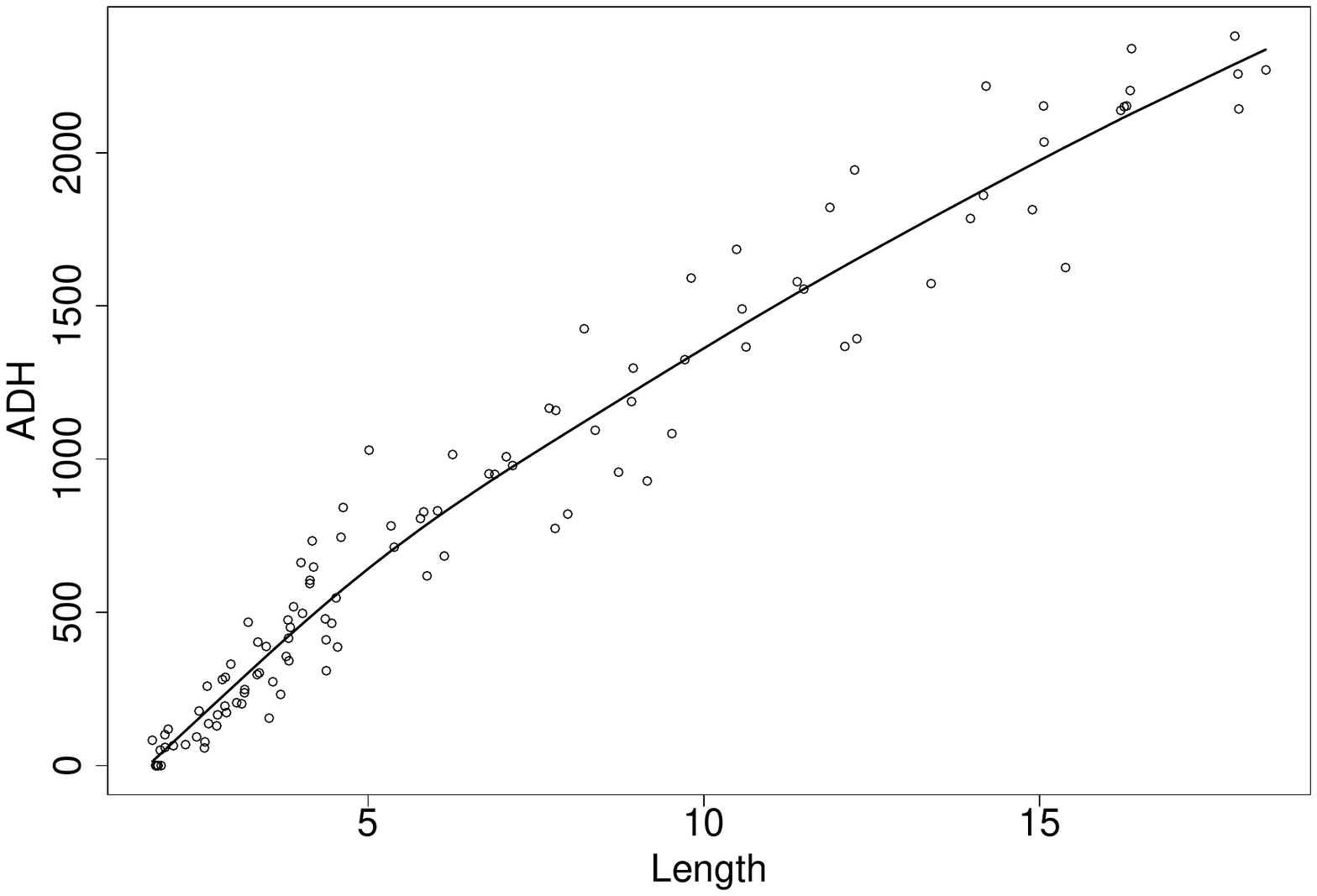}
	}
	\caption{\label{fig:sim4} Comparison between the proposed procedure based on growth curve reconstruction and the one based on  Accumulated Degrees Hours (ADH) in the considered simulated scenarios.  (a): Boxplots of the estimated hatching times for method based on ADH (white) and the  the proposed method based on growth curve reconstruction (gray) for simulated scenario (b) described in the text (varying temperature profile observed with random noise), for two values of $\sigma$, the standard deviation of the additive noise. The horizontal line denotes the true hatching time. (b): Scatterplot of experimental data for larval length and ADH and estimated relationship via smoothing splines (solid black line). }
\end{figure}

We then compare our proposed methods with existing techniques in these simulated scenarios. While the most commonly used approach based on the accumulated degrees hours (ADH) model is usually carried out by leaving the larvae in an incubator until pupation, we could use as benchmark a relationship between ADH and length estimated from the experimental data. We use experimental data on \emph{Calliphora vomitoria} growth in \citet{richards2017} to estimate this relationship nonparametrically via smoothing splines using the \texttt{gam} function from the R package \texttt{mgcv}. Figure \ref{fig:sim4}(b) shows the raw data and the estimated curve. We then use this estimated curve to predict the hatching time associated with the ADH predicted in correspondence of the average observed larval length. We set a larval development threshold (i.e. the temperature below which there is no growth in the larvae) to $4.3$ degrees Celsius, as observed experimentally for the larval development phase up to pupariation. We consider here only the more realistic scenario (b) of varying temperature profile and the results are shown in Figure \ref{fig:sim4}(a). The proposed method based on the reconstruction of the growth curve appears to be more accurate than the one based on the ADH curve. Although the ADH estimation can be better tuned by changing the lower developmental threshold (and potentially having it non constant along the larval development), this is often challenging in practice and it would require to develop a model for ADH similar to the one we propose here for the growth curve.

\subsection{First case study}
\begin{figure}[h!]
	\centering
	\includegraphics[width=0.45\textwidth, height = 6cm]{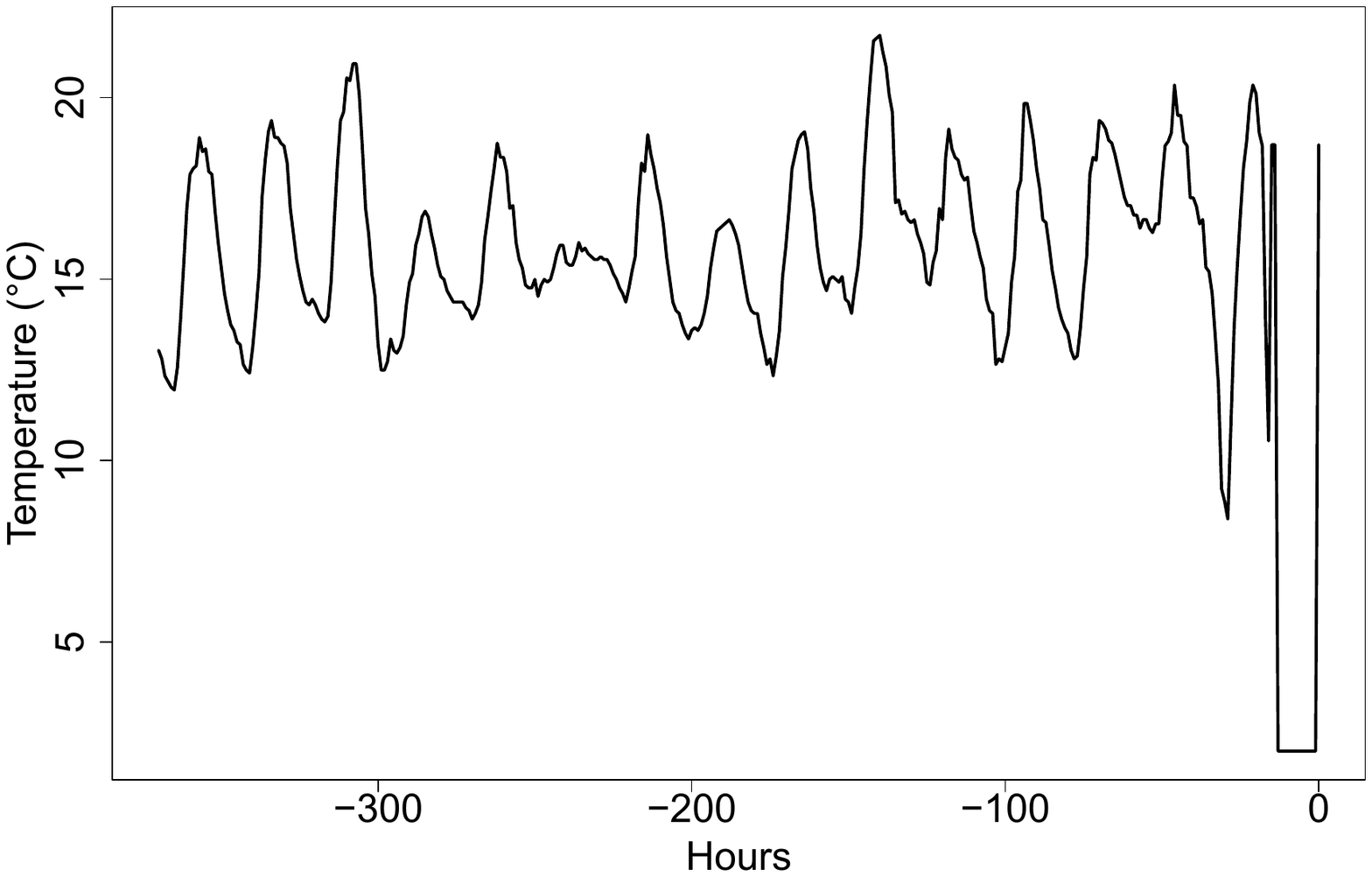}
	\includegraphics[width=0.45\textwidth, height = 6cm]{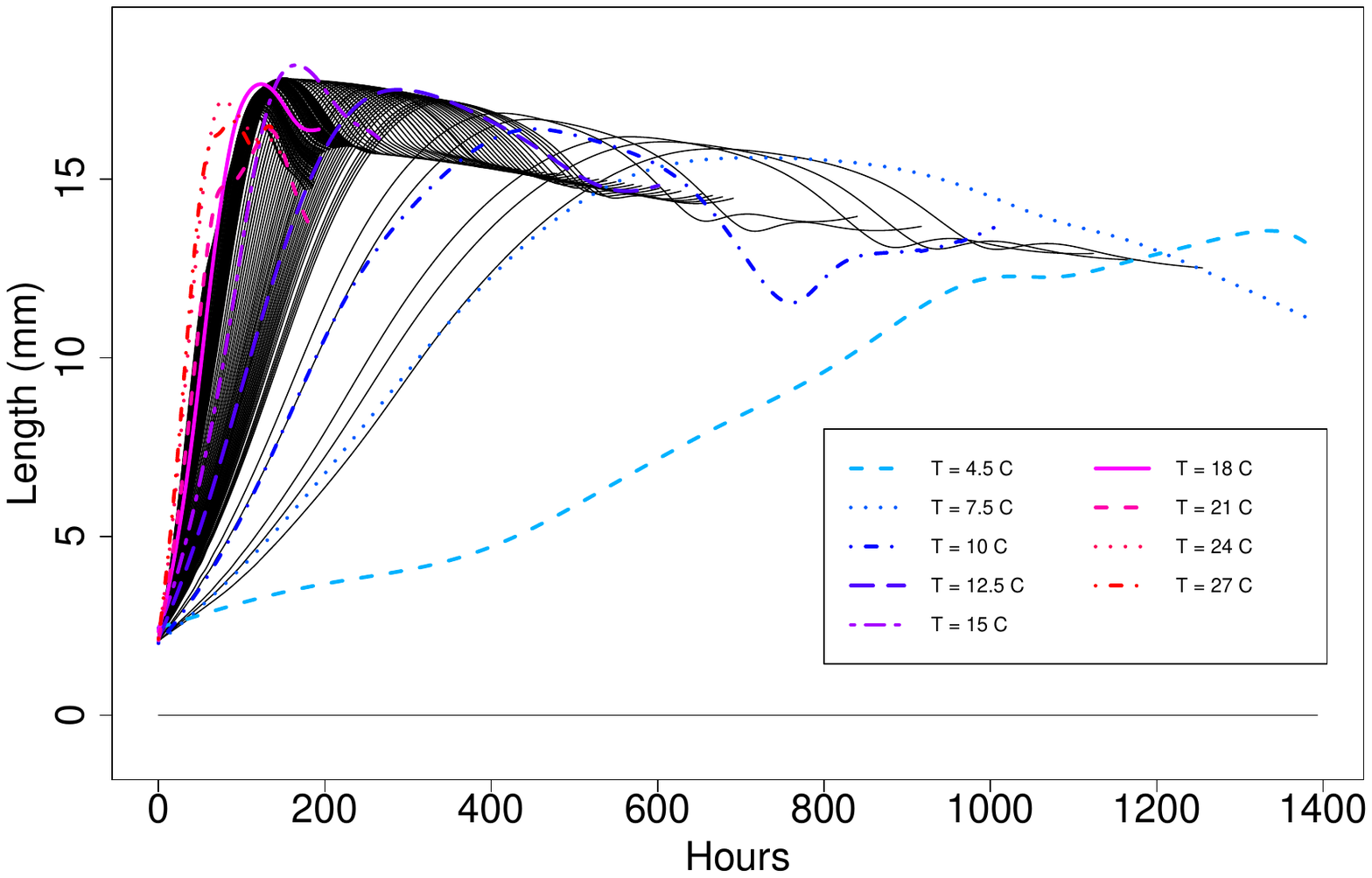}
	\caption{\label{fig:temp303} Left: Temperature profile to which the body was subjected in the hours before larval lengths were measured. Note that between the body discovery and the collection of the larvae the body was stored in a fridge for a few hours. Right: Estimated constant-temperature growth curves for each temperature in the observed interval (black lines), the temperature corresponding to observed experimental temperature are highlighted. Note the constant (zero) growth corresponding to the temperature in the fridge.}
\end{figure}

We consider a first case study where $n_{obs}=70$ \emph{Calliphora vicina} post-feeding larvae were collected from the body. In this investigation, there was not a unique crime scene since the body was moved between death and discovery. For this reason, the temperature profile to which the body was subjected is provided here by forensic experts based on the information about the body location coming from the investigation. Figure \ref{fig:temp303} illustrates the temperature time series for the 371 hours before the time the larvae were killed, prior to subsequent measurement, together with the constant temperature growth profiles corresponding to each observed temperature in the series. The interval of 371 hours was the largest one that was considered possible by forensic scientists at the scene. For this case, the application of the ADH method suggested as plausible interval for eggs' hatching the one between 275.5 and 208.5 hours before the measurements were taken \citep[][]{donovan2006larval}.

\begin{figure}[h!]
	\centering
	\subfigure[t][]{
		\centering
		\includegraphics[width=0.45\textwidth]{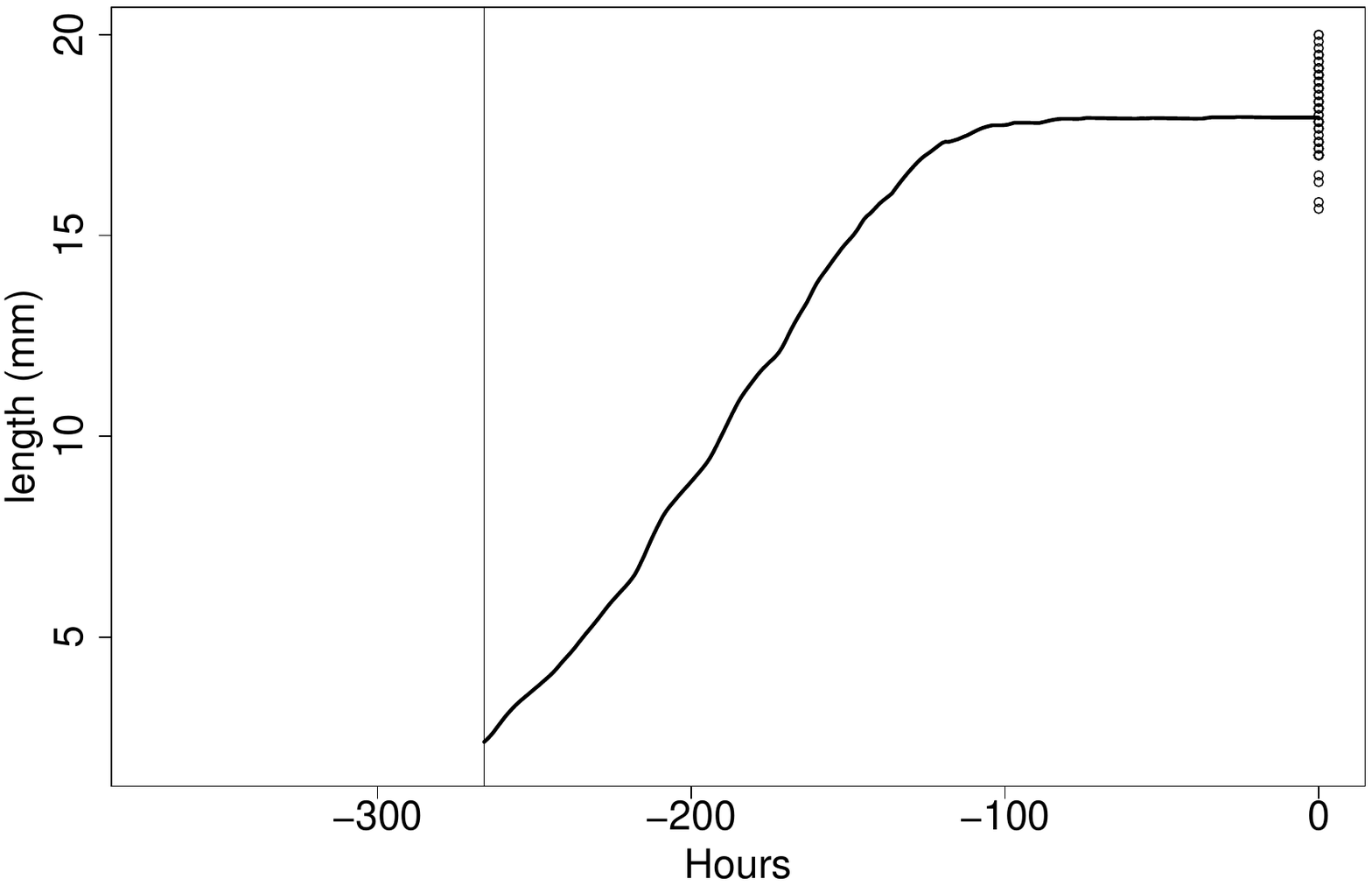}
	}
	\subfigure[t][]{
		\centering
		\includegraphics[width=0.45\textwidth]{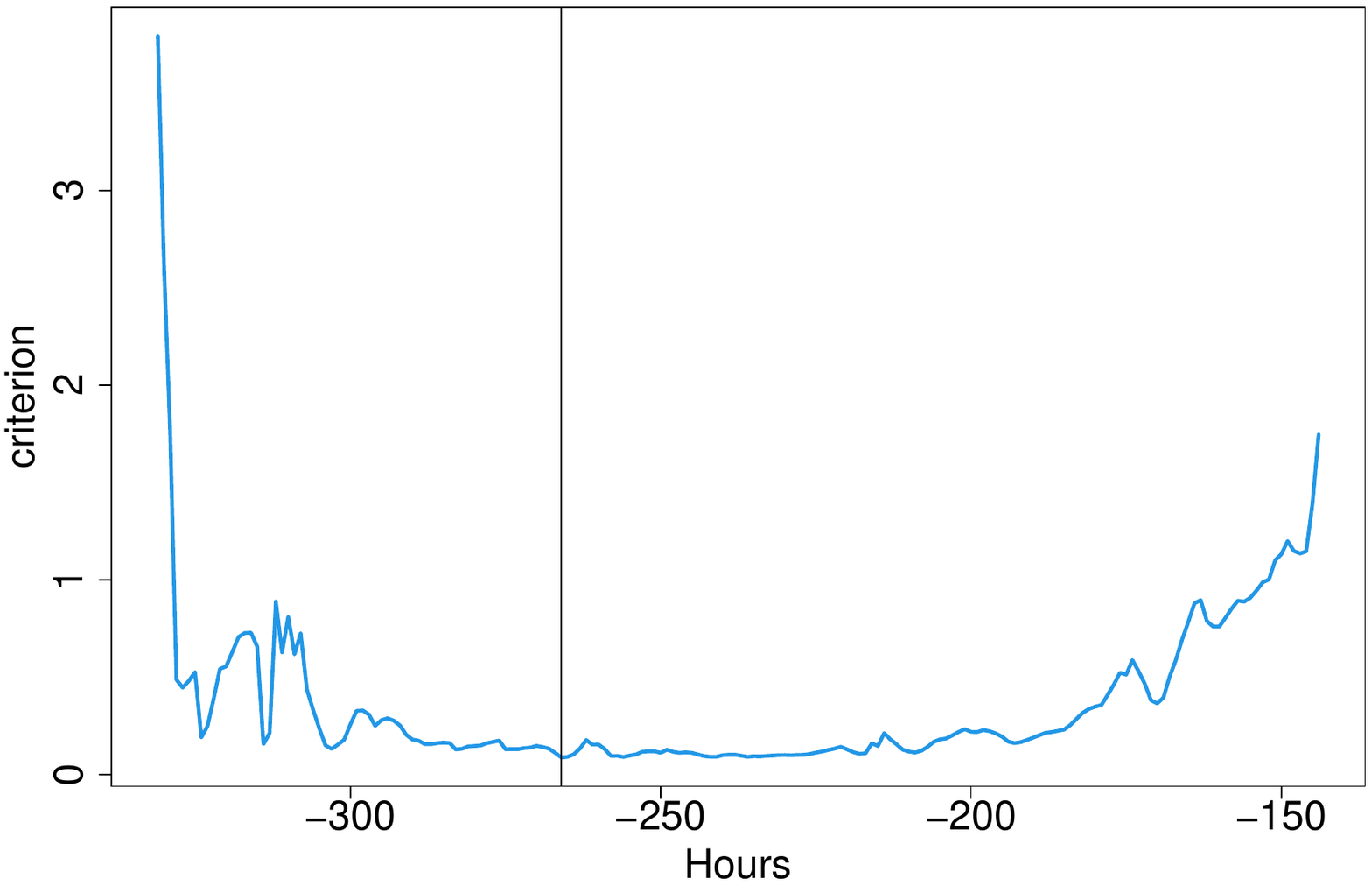}
	}
	\subfigure[t][]{
		\centering
		\includegraphics[width=0.45\textwidth]{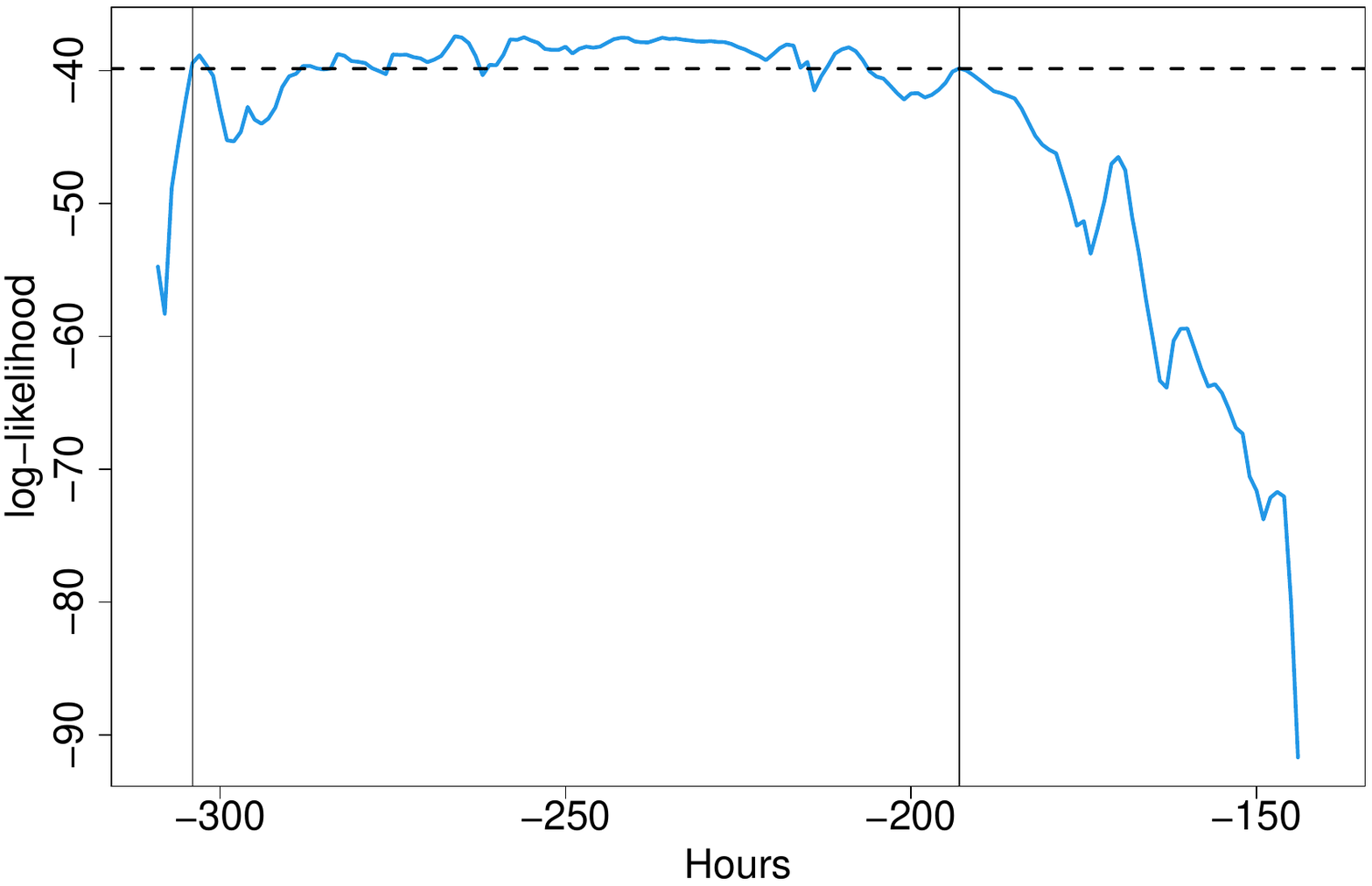}
	}
	\caption{\label{fig:est303} (a): Estimated growth curves for the observed larvae of \emph{Calliphora vicina}. The most likely hatching time is $266$ hours before the measurements were taken. (b): Profile of the criterion to be minimized as a function of hatching time.  (c): Log-likelihood in the region of the estimated hatching time and boundaries of the approximated $95\%$ confidence interval. }
\end{figure}

Figure {\ref{fig:est303} shows the growth curves for the hatching time estimated with (\ref{eq:crit}) and the profile of the objective function in the minimisation problem. The estimated hatching time is -266 hours (before the larval measurement), which is within the range obtained from the ADH method. However, the flat plateau in the criterion suggests little stability for the estimate. Indeed, if we assume a Gaussian distribution for the measurement errors, the $95\%$ approximated confidence interval procedure gives us an interval of $[-304,-193]$, which includes the range suggested by the ADH method. Note that here the computation of the criterion is restricted to the admissible region of hatching times for which the expected growth curve would have reached the post-feeding stage by the time the larvae were collected.
	
	In conclusion, the estimate for the hatching time provided by the proposed method is in agreement with the one from the ADH method in this case study, but the estimated uncertainty is larger than the bounds provided by accepted procedures based on ADH, for which no rigorous assessment of the uncertainty is available. Also, the fact that the point estimation of hatching time based on the proposed method gives us a larger time interval than the one obtained from the ADH method is consistent with the simulation results in Figure \ref{fig:sim4}, where the ADH method underestimated this time interval.

	\subsection{Second case study}
	\label{sec:case2}
	This second case study also includes the estimation of the temperature profile. A logger measures the temperature at the crime scene \emph{after} the body is discovered and this is compared to the data from the closest weather station, which are then used to estimate the past temperature at the crime scene. In particular, we are going to use a local polynomial kernel regression to estimate the crime scene temperature. The temperature at the crime scene, the weather station temperature measurements and the estimated crime scene temperature can be seen in Figure \ref{fig:temp307}, together with the constant temperature growth curves for all the estimated temperature values.
	
	\begin{figure}[h!]
		\centering
		\includegraphics[width=0.47\textwidth]{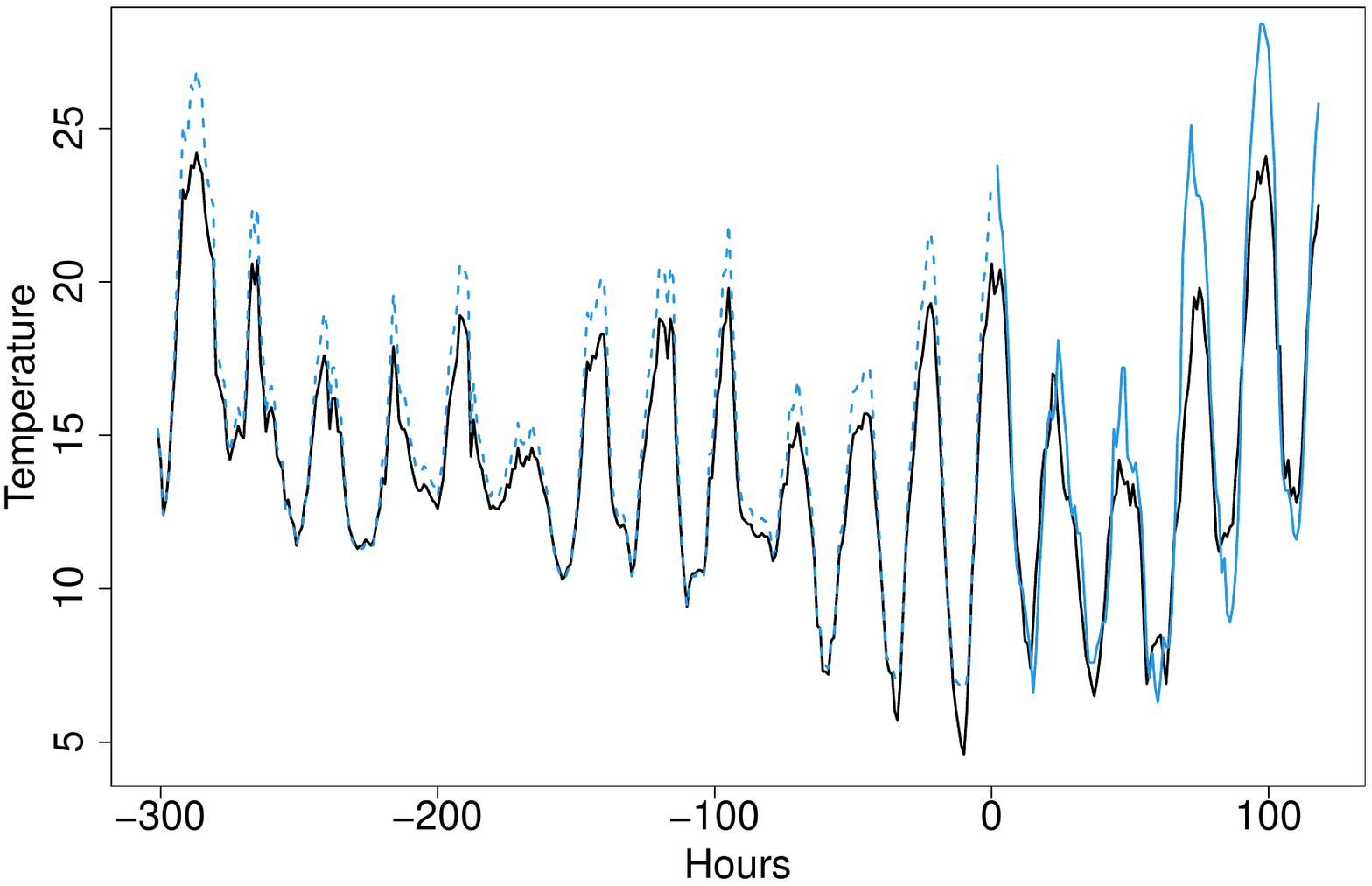}
		\includegraphics[width=0.44\textwidth]{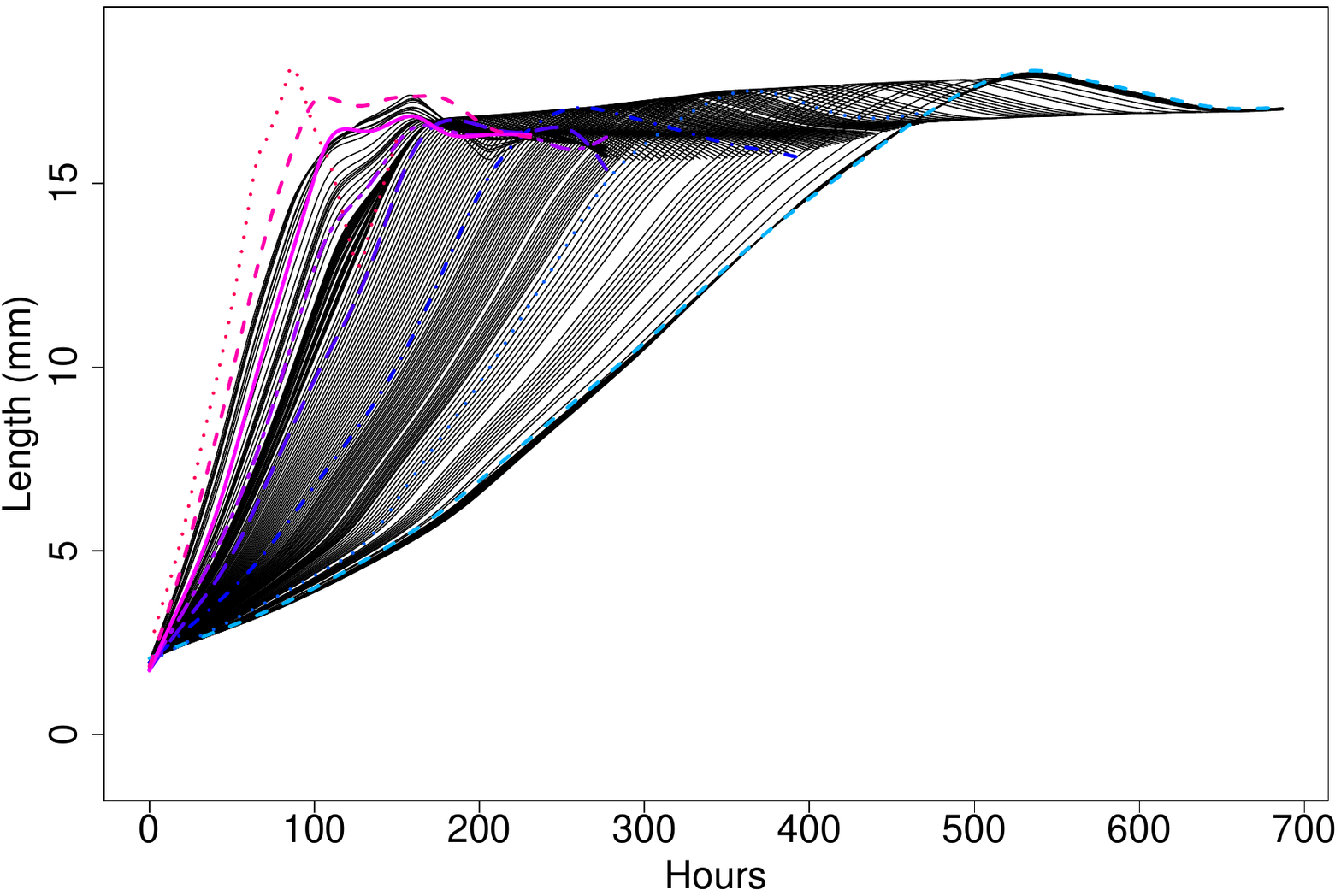}
		\caption{\label{fig:temp307} Left: Time series of temperature measured at the weather station closest to the crime scene (solid black line), temperature measured at the crime scene after body discovery (solid blue line) and estimated temperatures at the crime scene before body discovery (dashed blue line).  Right: Estimated constant-temperature growth curves for each temperature in the observed interval.}
	\end{figure}
	
	\begin{figure}[h!]
		\centering
		\subfigure[t][]{
			\centering
			\includegraphics[width=0.45\textwidth]{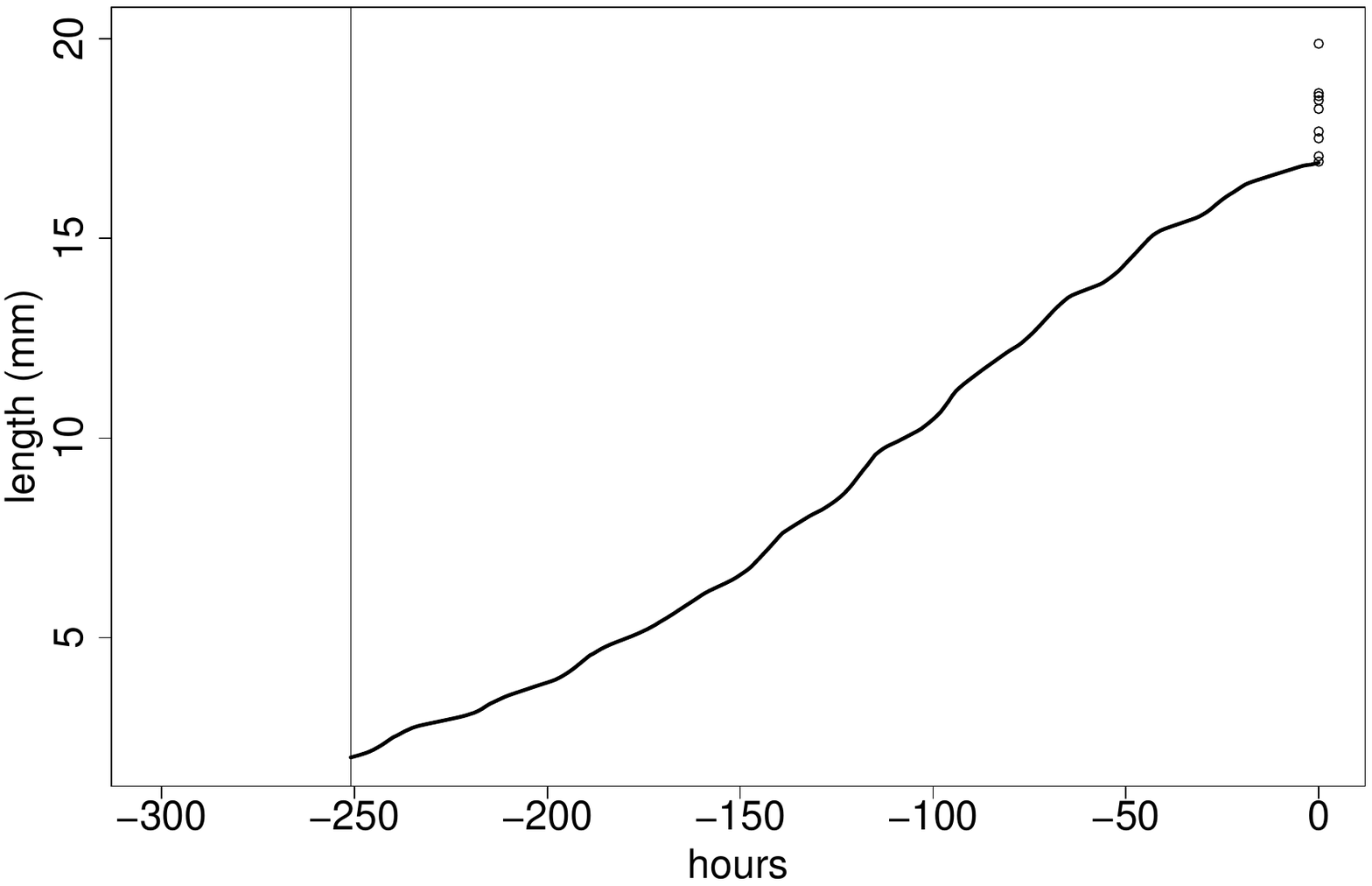}
		}
		\subfigure[t][]{
			\centering
			\includegraphics[width=0.45\textwidth]{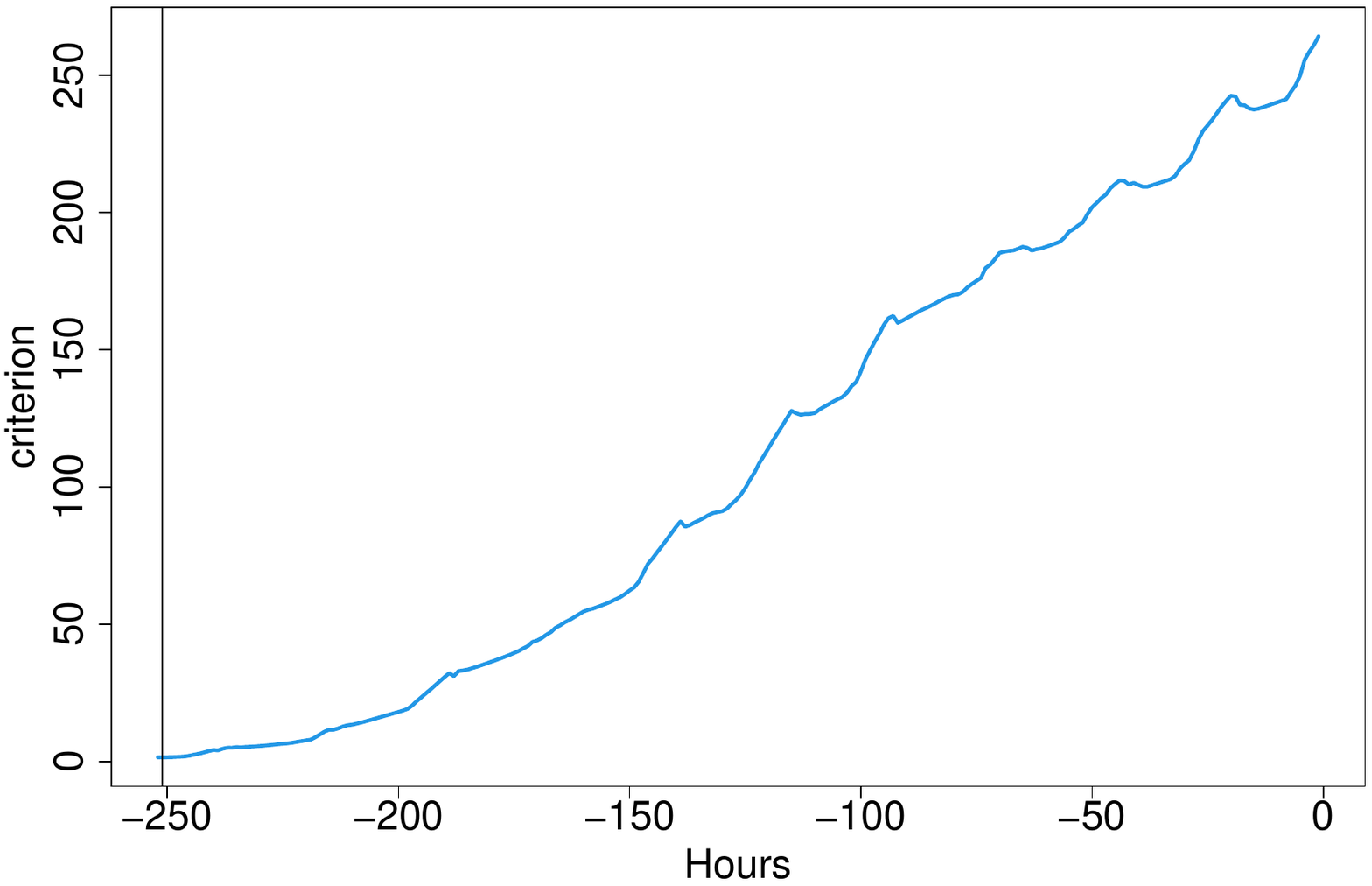}
		}
		\subfigure[t][]{
			\centering
			\includegraphics[width=0.45\textwidth]{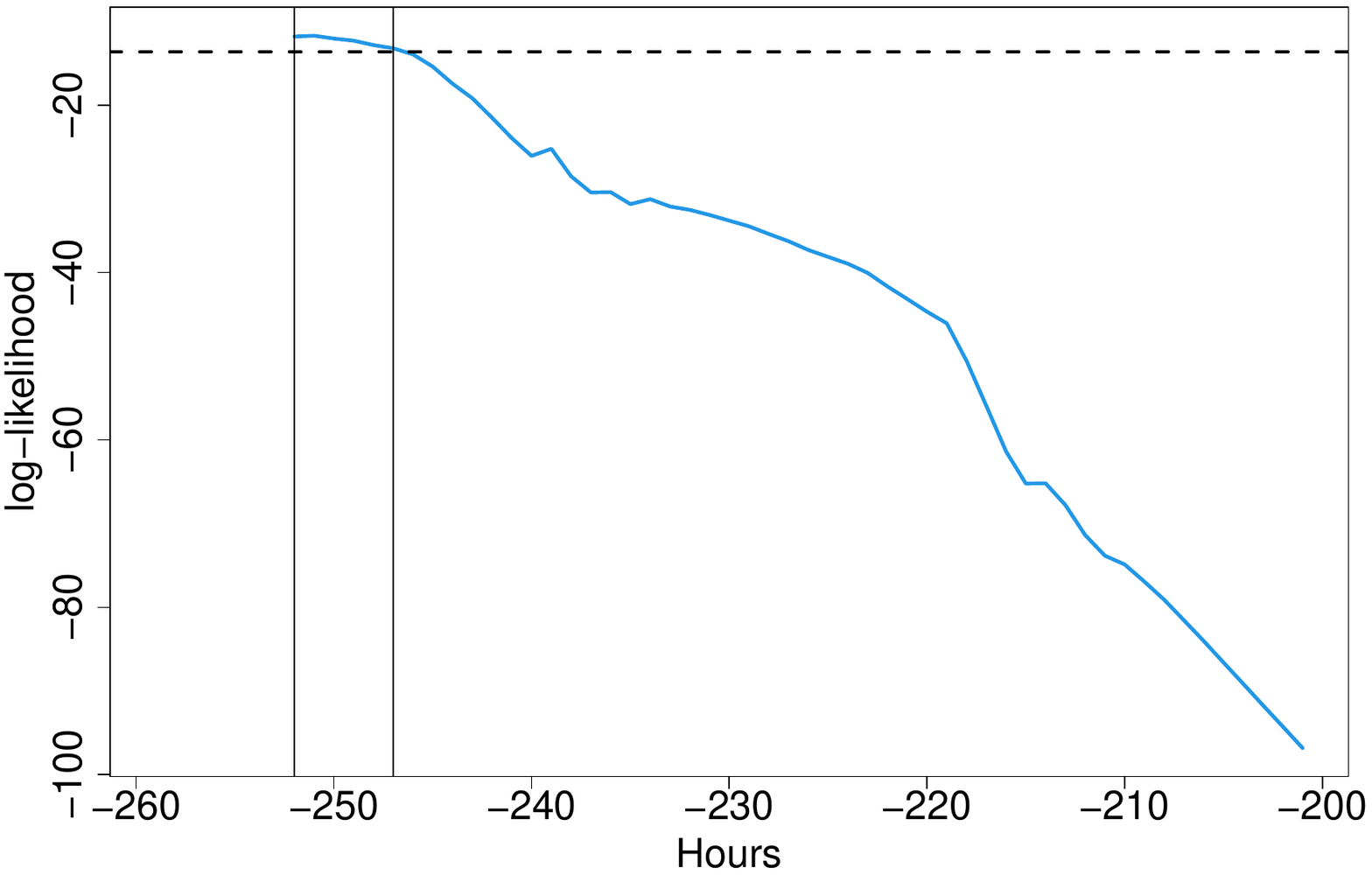}
		}
		\caption{\label{fig:est307} (a): Estimated growth curves for the observed larvae of \emph{Calliphora vomitoria}. The most likely hatching time is $251$ hours before the measurements were taken. (b): Profile of the criterion to be minimized as a function of hatching time. (c): Log-likelihood in the region of the estimated hatching time and boundaries of the approximated $95\%$ confidence interval.}
	\end{figure}

	At the crime scene, the lengths of $9$ \emph{Calliphora vomitoria} larvae still in the feeding phase were measured (among specimens of other species, for which experimental developmental data are not currently available). We are using here the development data on \emph{Calliphora vomitoria} collected at the Natural History Museum, London \citep{richards2017}. The assessment from forensic scientists, based on the application of the ADH method as well as qualitative considerations about the other species present at the scene, is that the body had been infested by bowfly eggs between $270$ and $240$ hours before the body was discovered. 
	
	Figure \ref{fig:est307} shows the estimated growth curve and the profile of the objective function. The estimate for the hatching time is $-251$ hours before the body discovery and the $95\%$ approximate confidence interval for a Gaussian error model is $[-252,-247]$. However, note that $-252$ is on the boundary of the admissible region and this makes the confidence interval based on the likelihood ratio statistics invalid. Moreover, we can see that the final expected length is too low with respect to the data and indeed the minimum for the criterion is reached at the boundary of the admissible region.

	This may suggest that the actual temperature at the crime scene was higher than predicted or that larvae reared in experimental cultures do not reach the potential maximum size reached in natural conditions. This is, of course, valuable information for the forensic scientists, unavailable from the ADH model.

	The original forensic reports proposed a time immediately adjacent to the time of the victim disappearance. This was because information from the investigation about the last time the victim has been seen alive was also taken into account. We can do the same with our model by using a Bayesian approach when the prior is taken to be a uniform between the earliest possible hatching time after the last sighting of the victim alive and the time of discovery of the body. The results can be seen in in Figure \ref{fig:bayes307}. Our method suggests strongly the earliest possible time for the hatching, in agreement with experts' judgement. It should be noted again that the growth curve does not appear to be able to reach the observed lengths.
	
	In conclusion, the experimental population did not develop to the average size of the larvae observed at the crime scene. This may be either because the development of larvae at the crime scene were affected by something that was not considered in the laboratory experiments (for example, larval-generated heat due to high concentration of specimens) or because the temperature at the crime scene was higher then expected during a portion of the development period. However, one useful advantage of the method over the ADH approach is that the estimated growth curve provides a diagnostic tool to spot any issue with the assumptions on the growth process.

	\begin{figure}[t!]
		\centering
		\includegraphics[width=0.6\textwidth]{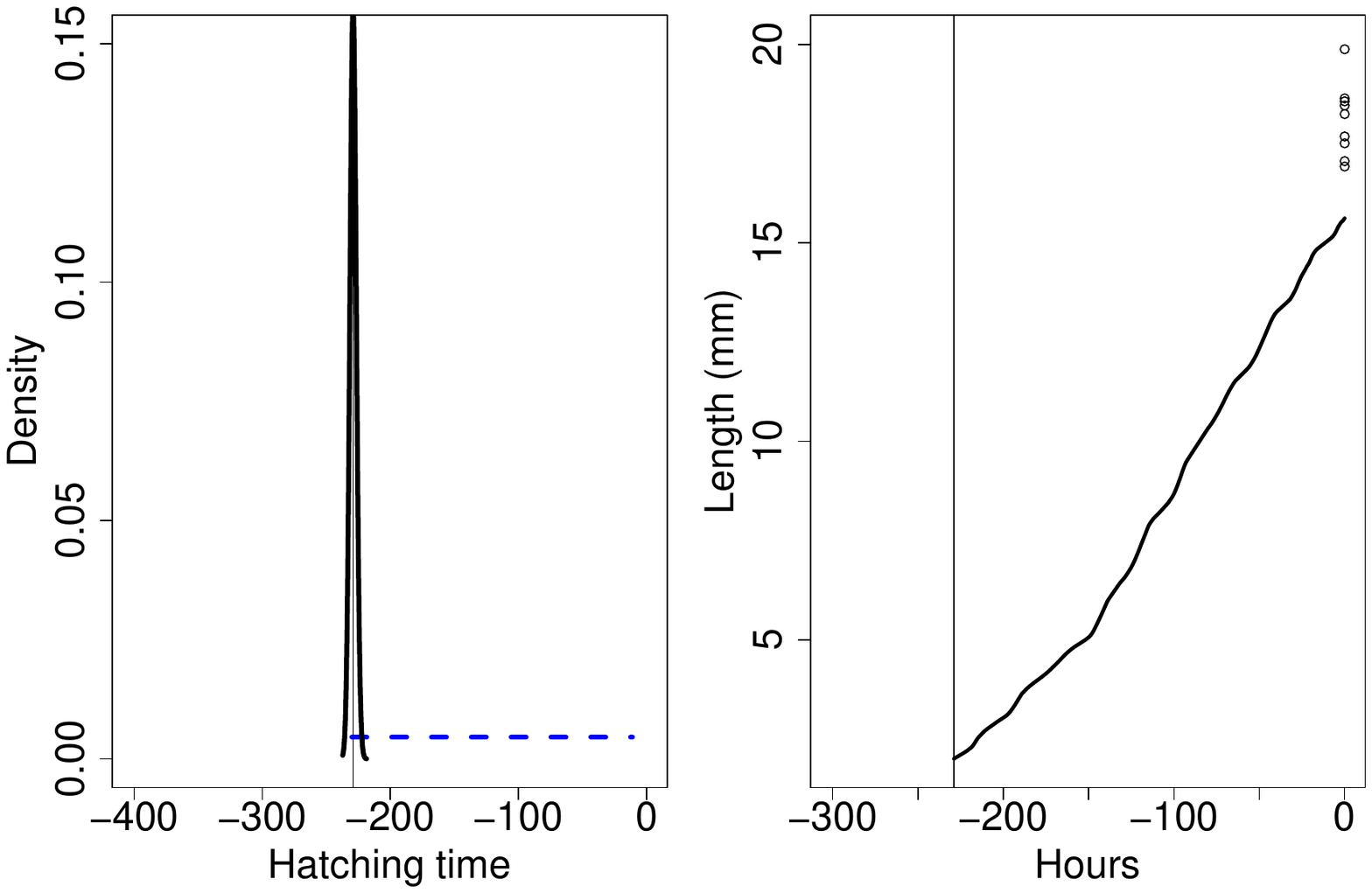}
		\caption{\label{fig:bayes307} Left: Prior (Blue dashed line, uniform distribution between the earliest possible hatching time after the last sighting of the victim alive and the time of discovery of the body) and posterior (black solid line) distributions of the hatching time for the larvae in the second case study. The vertical lines denote the maximum a posteriori (MAP) estimates of the hatching time. Right: Estimated expected growth curve from MAP estimates of the hatching times.}
	\end{figure}
	
	\section{Asymptotic properties}
	\label{sec:asymp}
	The main goal of this section is to state the asymptotic properties of the estimated hatching time, giving theoretical justification for our estimators, in addition to the empirical assessment given in the previous section. To this end, the estimating procedure is decomposed into four steps. In the first step the constant temperature growth profile $L_{T_k}(\cdot)$ and its derivative $L'_{T_k}(\cdot)$ are estimated from observed growth data set at a given ambient constant temperature $T_k$ belonging to the fixed design $T_1,\ldots,T_K$.  A decomposition of the estimated constant temperature growth profile $L_{T_k}(\cdot)$ into growth shape $S_{T_k}(\cdot)$ and warping function $w_{T_k}(\cdot)$ is achieved in step 2, at a given ambient temperature $T_k$ taken in the same fixed design. Step 3 focuses on the estimation of the constant temperature growth-profiles $L_T(\cdot)$ and its derivative $L'_T(\cdot)$ at any ambient temperature $T$ (i.e. $T$ may be outside the experimental design $T_1,\ldots,T_K$). In the last step the asymptotic behaviour of the estimated hatching date $\widehat{t}_h$ is provided. Assumptions are formally stated in the appendix and proofs can be found in the Supplementary Material. In order to make the theoretical developments more readable and accessible, the framework is voluntarily restricted to fixed designs, non random temperatures and only one species of fly. However, the setting can, of course, be extended to random designs, random temperatures and several species of fly, subject to some straightforward adjustments. Because all regularity assumptions involving $L_T$, $S_T$, and $w_T$ are uniform over a compact set of temperatures, all asymptotic results are also uniform with respect to the temperatures. \\~\\
	{\em Step 1: Estimating constant temperature growth profiles}.\\
	For a given constant temperature $T_k\in \{T_1, T_2, \ldots, T_K \}$ one observes repeated larvae growth lengths $\left\{Y_{kjl};\, j=1,\ldots,n_k, \, l=1,\ldots, N_{kj} \right\}$ at  a grid of $n_k$ times after hatching $0=t^k_1 < t^k_2 < \cdots < t^k_{n_k} :=t_{pup, T_k}$. From this sample, one derives the growth profile $L_{T_k}(t)$ at constant temperature $T_k$ and time after hatching $t$ by means of the nonparametric regression model $Y_{kjl} \, = \, L_{T_k}(t^k_j) + \varepsilon_{kjl}$ and the local linear regression estimator
	\begin{equation} \label{eq:llr}
		( \widehat{a}, \widehat{b}) \, = \, \arg \min_{a,b} \sum_{j = 1}^{n_k} \omega_k(t^k_j) \left\{\overline{Y}_{kj} - a - b(t^k_j - t)  \right\}^2 K\left\{h_L^{-1}(t^k_j - t)\right\},
	\end{equation}
	with
	$$
	\widetilde{L}_{T_k}(t)\, = \, \widehat{a} \, = \, (1,0)^T \left(\bX_t^T \, \bK_t \, \bX_t \right)^{-1} \bX_t^T \, \bK_t \,  \overline{\bY^k},
	$$
	where, for any $k$, $\displaystyle \overline{Y}_{kj}:= \frac{1}{N_{kj}} \sum_{l=1}^{N_{kj}} Y_{kjl}$, $\displaystyle \overline{\bY^k} \, := \, \left( \overline{Y}_{k1},\ldots,  \overline{Y}_{k n_k} \right)^T \, \in \R^{n_k}$, the $(n_k \times 2)$ matrix $\displaystyle \bX_t \, = \, \left(
	\begin{array}{cccc}
		1 & 1 & \cdots & 1\\
		t^k_1-t & t^k_2-t & \cdots & t^k_{n_k}-t
	\end{array}
	\right)^T$, and the $(n_k \times n_k)$ diagonal matrix \linebreak $\displaystyle \bK_t \, := \, \mbox{diag}\left(  \omega_k(t^k_1)\, K\left\{h_L^{-1}(t^k_1 - t)\right\}, \ldots,  \omega_k(t^k_{n_k})\, K\left\{h_L^{-1}(t^k_{n_k} - t)\right\}  \right)$. In practice, the function $\omega_k$ is set in such a way $\omega_k(t^k_j) = N_{kj}/\sum_j N_{kj}$. The expression of the estimator of $L'_{T_k}(t)$, the first derivative of $L_{T_k}(t)$, is given by:
	$$
	\widetilde{L}'_{T_k}(t)\, = \, \widehat{b} \, = \, (0,1)^T \left(\bX_t^T \, \bK_t \, \bX_t \right)^{-1} \bX_t^T \, \bK_t \,  \overline{\bY^k}.
	$$
	As it is shown in the next section, to derive the consistency for the whole estimating procedure, the uniform rate of  convergence for $\widetilde{L}_{T_k}$ and its derivative $\widetilde{L'}_{T_k}$ are needed.
	
	\begin{theorem} \label{th1} Set $r_n \, = \, $ Under (\prettyref{hypo:growthprofile}), (\prettyref{hypo:kernel})-(\prettyref{hypo:samplesizesample2}), for any $T_k\in \left\{T_1, \ldots, T_K \right\}$ it holds:
		\begin{equation}\label{eq:length}
			\left\| \widetilde{L}_{T_k} - L_{T_k} \right\|_\infty  \, = \, O(h^2_L) + O_P\left\{ (n h^3_L)^{-1/3} \right\},
		\end{equation}
		and
		\begin{equation}\label{eq:lengthderiv}
			\left\| \widetilde{L'}_{T_k} - L'_{T_k} \right\|_\infty  \, = \, O(h^2_L) + O_P\left\{ (n h^6_L)^{-1/3} \right\}.
		\end{equation}
	\end{theorem}
	Although the literature on local linear regression is dense, the proof of (\ref{eq:length}) and (\ref{eq:lengthderiv}) needs some adjustments that are presented in the Supplementary Material.\\~\\
	{\em Step 2: Decomposing growth profiles into growth shape and warping functions}.\\
	For the growth profile $L_{T_k}(\cdot)$ at a given constant temperature $T_k$, one postulates the existence and unicity of a growth shape $S_{T_k}$ mapping $[0,1]$ into $\R$ and a strictly increasing warping function $w_{T_k}$ mapping $[0,1]$ into $[0,1]$ such that, for any time $t$ after hatching,  $L_{T_k}(t) \, = \, S_{T_k} \circ w_{T_k}(t/t_{pup,T}),$ where $S_{T_k}$ and $w_{T_k}$ are derived from $L_{T_k}(\cdot)$ by aligning the curves at 0, $t^k_{max}/t_{pup,T}$ and 1. For a given real $\alpha\in(0,1)$ and for any $k$ varying from 1 to $K$, the warping function $w_{T_k}$ is assumed to be a quadratic polynomial such that:
	$$
	w_{T_k}(0) \, = \, 0, \; w_{T_k}(t^k_{max}/t_{pup,T}) \, = \, \alpha, \; w_{T_k}(1) \, = \, 1.
	$$
	In other words, the warping function $w_{T_k}$ is a strictly increasing function interpolating the three points with coordinates $(0;\,0)$, $(t^k_{max}/t_{pup,T};\,\alpha)$, and $(1;\,1)$. Imposing $w_{T_k}$ in the space of quadratic polynomials ensures its unicity. This procedure is commonly called registration (see Ramsay and Silverman, 2005).
	So, our aim is to state asymptotic properties of the $K$ estimated pairs $(\widetilde{S}_{T_1}, \widetilde{w}_{T_1}), \ldots, (\widetilde{S}_{T_K}, \widetilde{w}_{T_K})$  and their corresponding derivatives $(\widetilde{S}'_{T_1}, \widetilde{w}'_{T_1}), \ldots, (\widetilde{S}'_{T_K}, \widetilde{w}'_{T_K})$. For any $k\in \{1,\ldots, K\}$, let $\widetilde{t}^k_{max}$ be the time of maximum length derived from the estimated growth profile $\widetilde{L}_{T_k}$; the estimated warping function $\widetilde{w}_{T_k}$ is defined as the strictly increasing quadratic polynomial satisfying $\widetilde{w}_{T_k}(0)  =  0$, $\widetilde{w}_{T_k}(\widetilde{t}^k_{max}/t_{pup,T})  =  \alpha$, $\widetilde{w}_{T_k}(1)  =  1$, and, for any $u$ in $[0,1]$, the corresponding estimated growth shape $\widetilde{S}_{T_k}(u)= \widetilde{L}_{T_k} \left( t_{pup,T} \,\widetilde{w}^{-1}_{T_k}(u)\right)$, where $\widetilde{w}^{-1}_{T_k}$ is the reciprocal function of $\widetilde{w}_{T_k}$.
	
	\begin{theorem} \label{th2} Under (\prettyref{hypo:growthprofile})-(\prettyref{hypo:samplesizesample2}), for any $T_k\in \left\{T_1, \ldots, T_K \right\}$, $\widetilde{w}_{T_k}$ (resp. $\widetilde{S}_{T_k}$, $\widetilde{w}'_{T_k}$, $\widetilde{S}'_{T_k}$) converges to $w_{T_k}$ (resp. $S_{T_k}$, $w'_{T_k}$, $S'_{T_k}$) with the same rate of convergence $r_n \, = \, O(h^2_L) + O_P\left\{ (n h^6_L)^{-1/3} \right \}$.
	\end{theorem}
	
	\noindent {\em Step 3: Estimating constant temperature growth profiles and corresponding first derivatives at any ambient temperature}.\\
	In this step, the challenge is to estimate the constant temperature growth profile $L_T$ at any ambient temperature $T$ (i.e. not only in the discrete grid $T_1, \ldots, T_K$). Based on the $K$-samples  $(\widetilde{S}_{T_1}, T_1),\ldots,(\widetilde{S}_{T_K}, T_K)$ and $(\widetilde{w}_{T_1}, T_1)$,  $\ldots, (\widetilde{w}_{T_K}, T_K)$ one is able to derive an estimator $\widehat{S}_T$ (resp. $\widehat{w}_T$) of the growth shape $S_T$ (resp. $w_T$) for any ambient temperature $T$ (see (\ref{eq:shapeatanytemp}) and (\ref{eq:warpingatanytemp})). So, for any temperature $T$, it is easy to deduce an estimator of the growth profile $L_T$ by setting $\widehat{L}_T(t) = \widehat{S}_T \circ \widehat{w}_T(t/t_{pup,T})$ for any $t$ in $[0, t_{pup,T}]$.  Similarly, based on the $K$-sample  $(\widetilde{S}'_{T_1}, T_1), \ldots, (\widetilde{S}'_{T_K}, T_K)$ and $(\widetilde{w}'_{T_1}, T_1), \ldots, (\widetilde{w}'_{T_K}, T_K)$, one can derive the estimation of $S'_T$ and $w'_T$ at any temperature $T$ by setting
	$$
	\widehat{S}'_{T} \, = \, \frac{ \sum_{k=1}^K \widetilde{S}'_{T_k} \, K_{S'}\left\{h_{S'}^{-1}(T_k - T)\right\} }{ \sum_{k=1}^K K_{S'}\left\{h_{S'}^{-1}(T_k - T)\right\} } \textrm{ and }
	\widehat{w}'_{T} \, = \, \frac{ \sum_{k=1}^K \widetilde{w}'_{T_k} \, K_{w'}\left\{h_{w'}^{-1}(T_k - T)\right\} }{ \sum_{k=1}^K K_{w'}\left\{h_{w'}^{-1}(T_k - T)\right\} },
	$$
	where $K_{S'}$ (resp. $K_{w'}$) is also a kernel function and $h_{S'}$ (resp. $h_{w'}$) the non-negative smoothing parameter. Because 
	
	$$L'_T (t)\, = \, S'_T \left( w_T(t/t_{pup,T})\right) w'_T(t/t_{pup,T}) \, t_{pup,T}^{-1}, $$ for any temperature $T$, the derivative of the growth profile $L'_T(t)$ for any $t$ in $[0, \widehat{t}_{pup,T}]$ is estimated with $\widehat{L}'_T(t) \, = \, \widehat{S}'_T \left( \widehat{w}_T(t/\widehat{t}_{pup,T})\right) \widehat{w}'_T(t/\widehat{t}_{pup,T}) \, \widehat{t}_{pup,T}^{-1}$.
	
	\begin{theorem} \label{th3} Under (\prettyref{hypo:growthprofile})-(\prettyref{hypo:samplesizesample2}) and for any temperature $T$:
		\begin{eqnarray} \label{eq:growthprofileatanytemp} \nonumber
			\left\| \widehat{L}_{T} - L_T \right\|_\infty & = & O(h_{pup}) \, + \, O(h_S) \, + \, O(h_w) + \,  O_P\left(1/\sqrt{K \, h_{pup}}  \right) \, + \, r_n,
		\end{eqnarray}
		and
		\begin{eqnarray} \nonumber
			\left\| \widehat{L}'_T - L'_T \right \|_\infty & = &O(h_{pup})  \, + \, O(h_{S'}) \, + \, O(h_w)+ \, O(h_{w'}) + \, O_P\left\{ (K\, h_{pup})^{-1/2} \right \} \, + \, r_n.
			\label{eq:derivprofileatanytemp2}
		\end{eqnarray}
	\end{theorem}
	
	\noindent {\em Step 4: Estimating hatching time}.\\
	Before tackling the estimation of the unknown hatching date $t_h$, one focuses on the varying temperature growth length $L(t)$ at the time $t$ after hatching which is assumed to depend on the temperature process $\{T(v), \, v \in [0, t)  \}$. One postulates that the growth length $L(t)$ given the set of temperature variations $\{T(v), \, v \in [0, t)  \}$ satisfies the dynamic growth model:
	$$
	L(t) - L(0) \, = \, \int_0^t \left( \left. \frac{d L_{T(v)}(u) }{du}\right|_{u= L^{-1}_{T(v)}\left\{  L(v) \right\}}  \right) dv \, = \, \int_0^t \left\{ L'_{T(v)} \circ L^{-1}_{T(v)} \circ L(v)  \right\} dv,
	$$
	where $L(0)=L_{T(0)}(0)$ is the length at the hatching time. This formulation can be expressed in terms of differential equation: $L'(t)  \, = \,  L'_{T(t)} \circ L^{-1}_{T(t)} \circ L(t)$. In practice, one uses a fine grid of time $0=t_0<t_1<t_2<\cdots < t_{M-1}<t_M<t <t_{M+1}$ such that $\sup_\ell |t_{\ell+1}-t_\ell|=O(M^{-1})$ and the previous dynamic growth model can be approximated by its discretized version:
	$$
	L(t_M) - L(t_1) \, = \,  \sum_{\ell = 1}^M (t_{\ell+1} - t_\ell)\left\{ L'_{T(t_\ell)} \circ L^{-1}_{T(t_\ell)} \circ L(t_\ell)  \right\} + O(M^{-1}),
	$$
	with $L(t_1)=L_{T(t_1)}(t_1)$. So, the estimator $\widehat{L}$ of the varying temperature length profile $L$ is defined as follows:
	$$
	\widehat{L}(t_M) - L(t_1) \, = \,  \sum_{\ell = 1}^M (t_{\ell+1} - t_\ell)\left\{ \widehat{L}'_{T(t_\ell)} \circ \widehat{L}^{-1}_{T(t_\ell)} \circ \widehat{L}(t_\ell)  \right\}.
	$$
	Once the varying temperature growth profile is estimated, the ultimate task is to compute the hatching time and the hatching interval $H$. To this end, one observes on the crime scene at a given date $t^*$ a sample of i.i.d. larval lengths $Y^*_1, \ldots, Y^*_{n_{obs}}$ corresponding to the growth length $L(t^* - t_h)$ reached at time $t^* - t_h$ ($:= H$) after hatching and computed from the past outdoor temperature time series $\{ T(t^* - t);\, t\in [t_h,\, t^*] \}$, where $t_h$ stands for the unknown hatching date. Moreover, for $i=1, \ldots, n_{obs}$, one assumes $Y^*_i \, = \, L(t^* - t_h) + \epsilon_i$ where the $\epsilon_i$'s are zero mean random errors. Then, the estimated date of hatching $\widehat{t}_h$ is the minimizer of $\widehat{Q}(t):=\left( \widehat{L}(t^* - t) - \overline{Y}^*  \right)^2$, which gives the estimation of the hatching interval: $\widehat{H}=t^* - \widehat{t}_h$.
	
	\begin{theorem} \label{th4} Under (\prettyref{hypo:growthprofile})-(\prettyref{hypo:samplesizesample2}) it holds:
		\begin{eqnarray*}
			\widehat{t}_h - t_h & = &  O(h_{pup}) \, + \, O(h_{S}) \, + \, O(h_{S'}) \, + \, O(h_w) \, + \, O(h_{w'}) \\ \label{eq:length5}
			& & \, + \, O(M^{-1}) \, + \, O_P\left\{ (K\, h_{pup})^{-1/2} \right \} \, + \, O_P\left( n_{obs}^{-1/2}\right) \, + \, r_n.
		\end{eqnarray*}
	\end{theorem}
	As a by-product, one gets under same assumptions the consistency of the hatching interval:
	\begin{eqnarray*}
		\widehat{H} - H & = &  O(h_{pup}) \, + \, O(h_{S}) \, + \, O(h_{S'}) \, + \, O(h_w) \, + \, O(h_{w'}) \\ \label{eq:length5}
		& & \, + \, O(M^{-1}) \, + \, O_P\left\{ (K\, h_{pup})^{-1/2} \right \}  \, + \, O_P\left( n_{obs}^{-1/2}\right)  \, + \, r_n.
	\end{eqnarray*}

	\section{Conclusions}
	We have described a functional data approach to incorporate the information from the constant temperatures used to gather experimental data into the estimation of the crime scene varying temperature growth profiles. This can be used to estimate the most likely hatching time for the observed larvae. The proposed method has some advantages over the existing ADH approach. First of all, it can be applied directly to the lengths of the larvae measured at the crime scene, without the need to wait for the larvae to develop up to the next stage of the life cycle in incubators, as can be required by ADH methods based on duration of discrete life stages. Second, the proposed method allows forensic scientists to also consider the estimated growth curve as a model diagnostic tool to highlight any problematic situations, as seen in the second case study. Moreover, when larvae are observed in the middle of the development process, the method provides a more accurate estimate of the hatching time together with an estimate of the uncertainty. Note that this was not the case in the two case studies we considered, where larvae had already reached the maximum size or were in the post-feeding phase.
	
	We have demonstrated both theoretically and empirically that the functional data approach can provide good estimates for the interval from hatching time to discovery of the body. These are, of course, subject to a number of uncertainties, such as those relating to temperature estimation as well as the inherent biological variability in both the experimental and crime scene entomological data. We have shown that this uncertainty can be captured both in a frequentist or Bayesian setting, allowing the model to be used in either framework, depending on the availability of additional prior knowledge and the legal requirements of the court.
	
	It needs to be stressed that the methods by no means give a definitive conclusion about the post mortem interval on their own, since, on the one hand, many factors can delay the access of flies to the body \citep[see, e.g.,][]{bhadra2014factors} and, on the other hand, only expert judgement about the surroundings can guarantee that the observed larval specimens are the oldest to have colonised the body.  However, they do yield more complete estimates of the growth of the larvae compared with the simpler accumulated degree hour models, and therefore are of use in fieldwork situations. For the future, we also need to expand the model to include the egg and intra-puparial stages of immature blow fly development, so that the proposed methods extend up to the time of colonisation. Further work would also consider a more comprehensive handling of the uncertainty in the Bayesian framework to account for the uncertainty in the estimation of the average growth process conditional on the temperature profile, as well as potentially the uncertainty on the temperature profile itself. Finally, the selection of the optimal bandwidth in the non-parametric regression is also still an open problem, this potentially being different for shape and warping functions and pupariation time. 
	
	\section*{Acknowledgements}
	
	John Aston appreciates the support of UK Engineering and Physical Sciences Research Council grant EP/K021672/2, ``Functional Object Data Analysis and its Applications''. Anjali Mazumder appreciates the support of NIST and its funding of Center for Statistics and Applications in Forensic Evidence, and of the UKRI AI for Science and Government Programme. Martin Hall and Cameron Richards are grateful to Clare Rowlinson for technical assistance and to the Department for Business, Innovation and Skills (BIS, UK Government) for funding their work.
	
	\section*{Supplementary Material}
	Supplementary material with the proofs of the results in Section 4 is available upon request.
	
	\appendices
	
	\section{Assumptions} \label{sec:a}
	Let us first start with assumptions about the constant temperature growth profile $L_T$, the growth shape $S_T$, and the warping function $w_T$. uniformly over a compact set of temperatures:
	\begin{itemize}
		\item[\refstepcounter{hypothese} \label{hypo:growthprofile} (H\thehypothese)] For any $T$, $L_T$ is a four-times continuously differentiable function over $(0,\, t_{pup})$ with a nonnull second derivative in the neighbourhood of the maximum length time: there exists $C>0$ and $\delta>0$, such that, for all $t\in (t_{max}-\delta, \, t_{max}+\delta)$, $|L''_T(t)|>C$,
		\item[\refstepcounter{hypothese} \label{hypo:shapeprofile} (H\thehypothese)] For any $T$, $S_T$ is four-times continuously differentiable and the following uniform Lipschitz property hold: it exists $0<M<\infty$ such that, for any temperatures $\tau_1$ and $\tau_2$, $\displaystyle \| S_{\tau_1} - S_{\tau_2} \|_\infty \leq M \, |\tau_1 - \tau_2|$ and $\displaystyle \| S'_{\tau_1} - S'_{\tau_2} \|_\infty \leq M \, |\tau_1 - \tau_2|$,
		\item[\refstepcounter{hypothese} \label{hypo:warpingprofile} (H\thehypothese)] For simplicity and unicity purposes, $w_T$ is assumed to be a strictly increasing 2nd degree polynomial for any $T$ and has the uniform Lipschitz property:  it exists $0<M<\infty$ such that, for any temperatures $\tau_1$ and $\tau_2$, $\displaystyle \| w_{\tau_1} - w_{\tau_2} \|_\infty \leq M \, |\tau_1 - \tau_2|$ and $\displaystyle \| w'_{\tau_1} - w'_{\tau_2} \|_\infty \leq M \, |\tau_1 - \tau_2|$.
	\end{itemize}
	About the kernel functions. Let $\mathcal{K}$ stand for $K_L(\cdot)$, $K_S(\cdot)$, $K_{S'}(\cdot)$, $K_w(\cdot)$ and $K_{w'}(\cdot)$:
	\begin{itemize}
		\item[\refstepcounter{hypothese} \label{hypo:kernel} (H\thehypothese)] $\mathcal{K}$ is a symmetric bounded continuously differentiable kernel function on its support with $\mathcal{K}'$ bounded such that $supp(\mathcal{K})=(-1,1)$, $\int \mathcal{K}(u) \, du =1$.
	\end{itemize}
	For the remaining assumptions, remember that for a given constant temperature $T_k\in \{T_1, T_2, \ldots, T_K \}$ one observes repeated larvae growth lengths $\left\{Y_{kjl};\, j=1,\ldots,n_k, \, l=1,\ldots, N_{kj} \right\}$ at  a grid of $n_k$ times after hatching $0=t^k_1 < t^k_2 < \cdots < t^k_{n_k} :=t_{pup, T_k}$.	
	
	About the variability of the observed $Y_{kjl}$'s for any $k=1,2,\ldots,K$:
	\begin{itemize}
		\item[\refstepcounter{hypothese} \label{hypo:variance} (H\thehypothese)] Set $\sigma_k^2(t^k_j) := Var(Y_{kjl})$; $\sigma_k^2(\cdot)$ is an integrable and continuously differentiable function.
	\end{itemize}
	About the weighted function $\omega_k(\cdot)$'s involved in the linear local mean squared minimization problem \prettyref{eq:llr}:
	\begin{itemize}
		\item[\refstepcounter{hypothese} \label{hypo:weightedfunction} (H\thehypothese)] For any $k=1,\ldots,K$, $\omega_k(\cdot)$ is a twice continuously differentiable function.
	\end{itemize}
	About the sample sizes, grid size and the bandwidths used in the estimating procedure, one requests:
	\begin{itemize}
		\item[\refstepcounter{hypothese} \label{hypo:samplesizesample1} (H\thehypothese)] set $\displaystyle n:= \inf_k n_k$;   $n$ tends to infinity,  $h_L$ tends to zero with $n$, and $n \, h^6_L$ tends to infinity with $n$,
		\item[\refstepcounter{hypothese} \label{hypo:samplesizesample2} (H\thehypothese)]  the number $K$ of experimental temperatures tends to infinity with $n$; $h_{pup}$, $h_S$, $h_{S'}$, $h_w$, $h_{w'}$ tends to zero with $K$ and $K \, h_{pup}$  tends to infinity with $K$; the size $p$ of the grid where the generic temperature profile is sampled tends to infinity; the sample size $n_{obs}$ of larval lengths observed at the crime scene tends to infinity.
	\end{itemize}

	
	\bibliographystyle{plainnat}
	{\small 
		\bibliography{growth}

\begin{thebibliography}{30}
\providecommand{\natexlab}[1]{#1}
\providecommand{\url}[1]{\texttt{#1}}
\expandafter\ifx\csname urlstyle\endcsname\relax
  \providecommand{\doi}[1]{doi: #1}\else
  \providecommand{\doi}{doi: \begingroup \urlstyle{rm}\Url}\fi

\bibitem[Amendt et~al.(2007)Amendt, Campobasso, Gaudry, Reiter, LeBlanc, and
  Hall]{amendt2007best}
J.~Amendt, C.P. Campobasso, E.~Gaudry, C.~Reiter, H.N. LeBlanc, and M.J.R.
  Hall.
\newblock Best practice in forensic entomology—standards and guidelines.
\newblock \emph{International Journal of Legal Medicine}, 121\penalty0
  (2):\penalty0 90--104, 2007.

\bibitem[Bhadra et~al.(2014)Bhadra, Hart, and Hall]{bhadra2014factors}
P.~Bhadra, A.J. Hart, and M.J.R. Hall.
\newblock Factors affecting accessibility to blowflies of bodies disposed in
  suitcases.
\newblock \emph{Forensic Science International}, 239:\penalty0 62--72, 2014.

\bibitem[Burfield et~al.(2015)Burfield, Neumann, and
  Saunders]{burfield2015review}
Riley Burfield, Cedric Neumann, and Christopher~P Saunders.
\newblock Review and application of functional data analysis to chemical
  data—the example of the comparison, classification, and database search of
  forensic ink chromatograms.
\newblock \emph{Chemometrics and Intelligent Laboratory Systems}, 149:\penalty0
  97--106, 2015.

\bibitem[Dias et~al.(2013)Dias, Garcia, and Schmidt]{dias2013hierarchical}
Ronaldo Dias, Nancy~L Garcia, and Alexandra~M Schmidt.
\newblock A hierarchical model for aggregated functional data.
\newblock \emph{Technometrics}, 55\penalty0 (3):\penalty0 321--334, 2013.

\bibitem[Donovan et~al.(2006)Donovan, Hall, Turner, and
  Moncrieff]{donovan2006larval}
S.E. Donovan, M.J.R. Hall, B.D. Turner, and C.B. Moncrieff.
\newblock Larval growth rates of the blowfly, calliphora vicina, over a range
  of temperatures.
\newblock \emph{Medical and Veterinary Entomology}, 20\penalty0 (1):\penalty0
  106--114, 2006.

\bibitem[Fan(1992)]{fan1992}
J.~Fan.
\newblock Design-adaptive nonparametric regression.
\newblock \emph{Journal of the American Statistical Association}, 87\penalty0
  (420):\penalty0 998--1004, 1992.

\bibitem[Fan(1993)]{fan1993local}
J.~Fan.
\newblock Local linear regression smoothers and their minimax efficiencies.
\newblock \emph{The Annals of Statistics}, 21:\penalty0 196--216, 1993.

\bibitem[Fan and Gijbels(1992)]{fan1992variable}
J.~Fan and I.~Gijbels.
\newblock Variable bandwidth and local linear regression smoothers.
\newblock \emph{The Annals of Statistics}, 20:\penalty0 2008--2036, 1992.

\bibitem[Ferraty and Vieu(2006)]{ferraty2006nonparametric}
F.~Ferraty and P.~Vieu.
\newblock \emph{Nonparametric functional data analysis: theory and practice}.
\newblock Springer Science \& Business Media, 2006.

\bibitem[Ferraty et~al.(2011)Ferraty, Laksaci, Tadj, and
  Vieu]{ferraty2011kernel}
F.~Ferraty, A.~Laksaci, A.~Tadj, and P.~Vieu.
\newblock Kernel regression with functional response.
\newblock \emph{Electronic Journal of Statistics}, 5:\penalty0 159--171, 2011.

\bibitem[Greenberg(1991)]{greenberg1991flies}
B.~Greenberg.
\newblock Flies as forensic indicators.
\newblock \emph{Journal of Medical Entomology}, 28\penalty0 (5):\penalty0
  565--577, 1991.

\bibitem[Hofer et~al.(2017)Hofer, Hart, Mart{\'\i}n-Vega, and
  Hall]{hofer2017optimising}
Ines~MJ Hofer, Andrew~J Hart, Daniel Mart{\'\i}n-Vega, and Martin~JR Hall.
\newblock Optimising crime scene temperature collection for forensic entomology
  casework.
\newblock \emph{Forensic science international}, 270:\penalty0 129--138, 2017.

\bibitem[Horv{\'a}th and Kokoszka(2012)]{horvath2012inference}
L.~Horv{\'a}th and P.~Kokoszka.
\newblock \emph{Inference for functional data with applications}, volume 200.
\newblock Springer Science \& Business Media, 2012.

\bibitem[James et~al.(2007)]{james2007curve}
Gareth~M James et~al.
\newblock Curve alignment by moments.
\newblock \emph{Annals of Applied Statistics}, 1\penalty0 (2):\penalty0
  480--501, 2007.

\bibitem[Kingsolver et~al.(2004)Kingsolver, Ragland, and
  Shlichta]{kingsolver2004quantitative}
Joel~G Kingsolver, Gregory~J Ragland, and J~Gwen Shlichta.
\newblock Quantitative genetics of continuous reaction norms: thermal
  sensitivity of caterpillar growth rates.
\newblock \emph{Evolution}, 58\penalty0 (7):\penalty0 1521--1529, 2004.

\bibitem[Kneip and Gasser(1992)]{kneip1992statistical}
Alois Kneip and Theo Gasser.
\newblock Statistical tools to analyze data representing a sample of curves.
\newblock \emph{The Annals of Statistics}, pages 1266--1305, 1992.

\bibitem[Kneip and Ramsay(2008)]{kneip2008combining}
Alois Kneip and James~O Ramsay.
\newblock Combining registration and fitting for functional models.
\newblock \emph{Journal of the American Statistical Association}, 103\penalty0
  (483):\penalty0 1155--1165, 2008.

\bibitem[Maldonado et~al.(2002)Maldonado, Staniswalis, Irwin, and
  Byers]{maldonado2002similarity}
Yolanda~Mu{\~N}oz Maldonado, Joan~G Staniswalis, Louis~N Irwin, and Donna
  Byers.
\newblock A similarity analysis of curves.
\newblock \emph{Canadian Journal of Statistics}, 30\penalty0 (3):\penalty0
  373--381, 2002.

\bibitem[Marron et~al.(2014)Marron, Ramsay, Sangalli, and
  Srivastava]{marron2014statistics}
J.S. Marron, J.O. Ramsay, L.M. Sangalli, and A.~Srivastava.
\newblock Statistics of time warpings and phase variations.
\newblock \emph{Electronic Journal of Statistics}, 8\penalty0 (2):\penalty0
  1697--1702, 2014.

\bibitem[Mart{\'\i}n-Vega et~al.(2017)Mart{\'\i}n-Vega, Simonsen, and
  Hall]{martin2017looking}
D.~Mart{\'\i}n-Vega, T.J. Simonsen, and M.J.R. Hall.
\newblock Looking into the puparium: Micro-ct visualization of the internal
  morphological changes during metamorphosis of the blow fly, calliphora
  vicina, with the first quantitative analysis of organ development in
  cyclorrhaphous dipterans.
\newblock \emph{Journal of Morphology}, 278:\penalty0 629--651, 2017.

\bibitem[Ramsay and Li(1998)]{ramsay1998curve}
James~O Ramsay and Xiaochun Li.
\newblock Curve registration.
\newblock \emph{Journal of the Royal Statistical Society: Series B (Statistical
  Methodology)}, 60\penalty0 (2):\penalty0 351--363, 1998.

\bibitem[Ramsay and Silverman(2005)]{ramsay2005functional}
J.O. Ramsay and B.W. Silverman.
\newblock \emph{Functional Data Analysis}.
\newblock Springer, 2005.

\bibitem[Reibe and Madea(2010)]{reibe2010promptly}
S.~Reibe and B.~Madea.
\newblock How promptly do blowflies colonise fresh carcasses? a study comparing
  indoor with outdoor locations.
\newblock \emph{Forensic Science International}, 195\penalty0 (1):\penalty0
  52--57, 2010.

\bibitem[Richards et~al.(unpublished)Richards, Rowlinson, and
  Hall]{richards2017}
C.S. Richards, C.C. Rowlinson, and M.J.R. Hall.
\newblock First full developmental data set for calliphora vomitoria and a
  consideration for the accumulated degree hour development model.
\newblock unpublished.

\bibitem[Ruppert and Wand(1994)]{ruppert1994multivariate}
D.~Ruppert and M.~P. Wand.
\newblock Multivariate locally weighted least squares regression.
\newblock \emph{The Annals of Statistics}, 23:\penalty0 1346--1370, 1994.

\bibitem[Srivastava et~al.(2011)Srivastava, Wu, Kurtek, Klassen, and
  Marron]{srivastava2011registration}
Anuj Srivastava, Wei Wu, Sebastian Kurtek, Eric Klassen, and James~Stephen
  Marron.
\newblock Registration of functional data using fisher-rao metric.
\newblock \emph{arXiv preprint arXiv:1103.3817}, 2011.

\bibitem[Tomberlin and Benbow(2015)]{tomberlin2015forensic}
Jeffery~Keith Tomberlin and M~Eric Benbow.
\newblock \emph{Forensic entomology: international dimensions and frontiers}.
\newblock CRC Press, 2015.

\bibitem[Wang et~al.(1997)Wang, Gasser, et~al.]{wang1997alignment}
Kongming Wang, Theo Gasser, et~al.
\newblock Alignment of curves by dynamic time warping.
\newblock \emph{The annals of Statistics}, 25\penalty0 (3):\penalty0
  1251--1276, 1997.

\bibitem[Warren et~al.(2017)Warren, {Pulindu Ratnasekera}, Campbell, and
  Anderson]{WARREN2017205}
J.-A. Warren, T.~D. {Pulindu Ratnasekera}, D.~A. Campbell, and G.~S. Anderson.
\newblock Initial investigations of spectral measurements to estimate the time
  within stages of protophormia terraenovae (robineau-desvoidy) (diptera:
  Calliphoridae).
\newblock \emph{Forensic Science International}, 278:\penalty0 205--216, 2017.

\bibitem[Zwietering et~al.(1990)Zwietering, Jongenburger, Rombouts, and
  Van't~Riet]{zwietering1990modeling}
MH~Zwietering, Il~Jongenburger, FM~Rombouts, and K~Van't~Riet.
\newblock Modeling of the bacterial growth curve.
\newblock \emph{Appl. Environ. Microbiol.}, 56\penalty0 (6):\penalty0
  1875--1881, 1990.

\end{thebibliography}
	}

\end{document}